\documentclass[a4paper, 11pt]{article}
\usepackage{jheppub, bm}
\usepackage{graphicx}
\usepackage[dvipsnames]{xcolor}
\usepackage{amsmath}
\allowdisplaybreaks
\usepackage{amssymb}
\usepackage{braket}

\usepackage{psfrag,graphicx,epsfig,amsmath,amssymb,bm}
\usepackage{hyperref}
\usepackage{slashed}
\usepackage{subfig}
\usepackage{xcolor}
\usepackage[normalem]{ulem}

\def \md {\mathrm{d}}
\newcommand{\nn}{\nonumber}

\usepackage{lipsum}
\allowdisplaybreaks

\begin{document}

\preprint{CPTNP-2025-024}

\makeatletter
\def\@fpheader{~}
\makeatother

\title{\boldmath
Spin correlations and Bell nonlocality in $\Lambda\bar\Lambda$ pair production from $e^+e^-$ collisions with a thrust cut
\unboldmath}

\author[a,e]{Shi-Jia Lin,}

\author[a]{Ming-Jun Liu,}

\author[a,b,c]{Ding Yu Shao,}

\author[d]{and Shu-Yi Wei}

\affiliation[a]{Department of Physics and Center for Field Theory and Particle Physics, Fudan University, Shanghai 200438, China}

\affiliation[b]{Key Laboratory of Nuclear Physics and Ion-beam Application (MOE), Fudan University, Shanghai 200438, China}

\affiliation[c]{Shanghai Research Center for Theoretical Nuclear Physics, NSFC and Fudan University, Shanghai 200438, China}

\affiliation[d]{Institute of Frontier and Interdisciplinary Science, Key Laboratory of Particle Physics and Particle Irradiation (MOE), Shandong University, Qingdao, Shandong 266237, China}

\affiliation[e]{Department of Physics, University of Chicago, Chicago, IL 60637, USA}

\emailAdd{sjlin21@m.fudan.edu.cn}
\emailAdd{mjliu24@m.fudan.edu.cn}
\emailAdd{dingyu.shao@cern.ch}
\emailAdd{shuyi@sdu.edu.cn}


\abstract{We present a comprehensive theoretical study of spin correlations in $\Lambda\bar{\Lambda}$ production from $e^+e^-$ annihilation, providing the theoretical predictions for the Belle II experiment. Using soft-collinear effective theory, we perform the first resummation of large logarithms for the longitudinal ($C_{LL}$) and transverse ($C_{TT}$) spin correlations for events with a cut on the thrust variable. Our calculation achieves next-to-next-to-leading logarithmic accuracy and incorporates the determination of polarized fragmenting jet functions. This framework provides robust predictions with significantly reduced theoretical uncertainties compared to fixed-order parton model approaches. Furthermore, we establish a direct mapping between the experimentally accessible spin correlation, $C_{TT}$, and a testable CHSH-Bell inequality. This result reframes $C_{TT}$ as a quantitative probe of quantum decoherence, providing a novel tool to measure the degree of parton-level entanglement that survives the fragmentation and hadronization process.}

\maketitle

\section{Introduction}
\label{sec:intro}

The study of hyperon polarization has long served as a crucial laboratory for testing our understanding of spin dynamics in Quantum Chromodynamics (QCD). The unexpected discovery of large transverse polarization for $\Lambda$ hyperons in unpolarized proton-nucleus collisions in 1976 presented a significant puzzle~\cite{Bunce:1976yb, Heller:1983ia}. This observation was in stark contradiction to the prevailing theoretical expectation that such single-spin asymmetries should be highly suppressed in perturbative QCD~\cite{Kane:1978nd}. This long-standing challenge has since motivated decades of experimental and theoretical investigations, firmly establishing that the mechanisms governing the spin of hadrons are far more intricate than initially presumed~\cite{Liang:1997rt, Filippone:2001ux, DAlesio:2007bjf}.

A key advantage of studying the $\Lambda$ hyperon is its weak decay, $\Lambda \to p\pi^-$, which acts as a self-analyzing polarimeter, allowing the hyperon's spin state to be reconstructed from the angular distribution of its decay products. While much of the historical focus has been on single-spin asymmetries, the production of $\Lambda\bar{\Lambda}$ pairs in the clean environment of electron-positron collisions offers a unique opportunity to probe the spin transfer in fragmentation through spin correlations of two-particle systems \cite{Chen:1994ar, Zhang:2023ugf, Chen:2024qvx, Yang:2024kjn}. These observables, such as the longitudinal ($C_{LL}$) and transverse ($C_{TT}$) spin correlations, provide direct access to the polarized fragmentation functions (FFs) that encode the non-perturbative physics of how a quark's spin is transmitted to a final-state hadron~\cite{Boussarie:2023izj, Leader:2001nas}. Most recently, the STAR collaboration at RHIC has presented the first evidence for spin correlations in $\Lambda\bar{\Lambda}$ pairs produced within the same jet from proton-proton collisions, opening a new observational window on the hadronization process~\cite{STAR:2025njp}.

Despite their fundamental importance, experimental data on $C_{LL}$ and $C_{TT}$ in $e^+e^-\to\Lambda\bar{\Lambda}X$ production are scarce, and a complete theoretical framework for jet-based observables is lacking. While foundational theoretical work exists within parton model frameworks incorporating DGLAP evolution~\cite{Yang:2024kjn, Barone:2003fy}, a systematic treatment of QCD evolution effects under the kinematic constraints imposed by an event shape like thrust has remained an open question. To provide reliable predictions for upcoming measurements at Belle II~\cite{Belle-II:2018jsg, Accardi:2022oog}—especially those involving event shape cuts such as thrust ($T$) \cite{Belle:2008fdv, Belle:2011cur, Belle:2018ttu}—a more sophisticated approach is required~\cite{Procura:2009vm, Jain:2011xz, Procura:2014cba, Lustermans:2019plv, Kang:2020yqw, Makris:2020ltr, Boglione:2020auc, Gamberg:2021iat, Boglione:2021wov, Boglione:2023duo, Fang:2025dee}. In the dijet limit ($T \to 1$), fixed-order calculations are spoiled by large Sudakov logarithms, necessitating an all-order resummation. In this work, we employ soft-collinear effective theory (SCET)~\cite{Bauer:2000yr, Bauer:2001ct, Bauer:2001yt, Bauer:2002nz, Beneke:2002ph} to systematically perform this resummation to next-to-next-to-leading logarithmic (NNLL) accuracy. This thrust-resummation framework is built upon fragmenting jet functions (FJFs)~\cite{Procura:2009vm, Jain:2011xz}, which we extend to the domain of polarized hyperon production within a fully resummed calculation, representing a state-of-the-art tool for calculating spin observables in the dijet region.

Beyond precision QCD, the production of spin entangled particles at high-energy colliders provides a novel arena for exploring fundamental quantum phenomena \cite{Abel:1992kz}. Much of the recent experimental and theoretical focus has been on systems where spin information can be accessed through electroweak decays that occur before hadronization. This approach has been successfully applied to top-quark pairs at the LHC, leading to the first observations of entanglement at hadron colliders by the ATLAS and CMS collaborations~\cite{ATLAS:2023fsd, CMS:2024pts}, and has been proposed for numerous other systems \cite{Severi:2021cnj, Altakach:2022ywa, Aguilar-Saavedra:2022uye, Ehataht:2023zzt, Fabbrichesi:2024wcd, Barr:2022wyq, Ashby-Pickering:2022umy, Aguilar-Saavedra:2022wam, Fabbrichesi:2023cev, Du:2024sly, Fabbri:2023ncz, Han:2023fci, Bi:2023uop, Wu:2024asu, Wu:2024mtj, Pei:2025non, Afik:2025grr, Ruzi:2025jql, Goncalves:2025xer, Cheng:2025zcf, Morales:2023gow, Morales:2024jhj, Goncalves:2025qem, Dong:2023xiw}. A more challenging but equally fundamental frontier is the study of spin entanglement between light quarks and gluons, whose spin correlations must be accessed through the non-perturbative hadronization process itself~\cite{Guo:2024jch, Cheng:2025cuv, vonKuk:2025kbv, Qi:2025onf}. This transition from partons to hadrons inevitably induces decoherence~\cite{Bertlmann:2004yg, Carney:2017jut, Carney:2017oxp, Carney:2018ygh, Neuenfeld:2018fdw, Semenoff:2019dqe, Schlosshauer:2019ewh, Burgess:2024heo, Salcedo:2024smn, Salcedo:2024nex, Aoude:2025ovu}, making the final-state spin correlations a sensitive probe of the quantum dynamics of fragmentation.

Building on these developments, we investigate the spin correlations of the $\Lambda\bar{\Lambda}$ system from a quantum entanglement perspective. A key result of our work is the establishment of a direct mapping between the experimentally accessible spin correlation, $C_{TT}$, and a testable Clauser-Horne-Shimony-Holt (CHSH) Bell parameter. This connection reframes the measurement of $C_{TT}$ as a quantitative probe of the decoherence induced by the fragmentation and hadronization process, allowing us to ask: how much of the initial quantum entanglement survives the complex fragmentation dynamics of QCD?\footnote{We emphasize that our analysis is performed within the framework of quantum field theory and is aimed at quantifying the decoherence caused by QCD fragmentation and hadronization. We do not attempt to interpret collider measurements as fundamental tests of local hidden-variable theories, a subject of recent debate in the Refs.~\cite{Li:2024luk, Bechtle:2025ugc, Abel:2025skj, Low:2025aqq}.}

Our paper is organized as follows. We begin in Sec.~\ref{sec:parton_model} by reviewing the baseline parton model calculation of the spin correlations. In Sec.~\ref{sec:thrust}, we develop our primary theoretical framework, deriving the thrust-resummed cross section at NNLL accuracy using SCET and polarized FJFs. We then explore the connection between these QCD observables and quantum information in Sec.~\ref{sec:decoherence}, linking the spin correlations to Bell variables. Our numerical predictions are presented in Sec.~\ref{sec:Pheno}. Finally, we summarize our findings and discuss future prospects in Sec.~\ref{sec:Sum}.

\section{Parton model calculation}\label{sec:parton_model}

\begin{figure}[t]
    \centering
    \includegraphics[scale = 0.4]{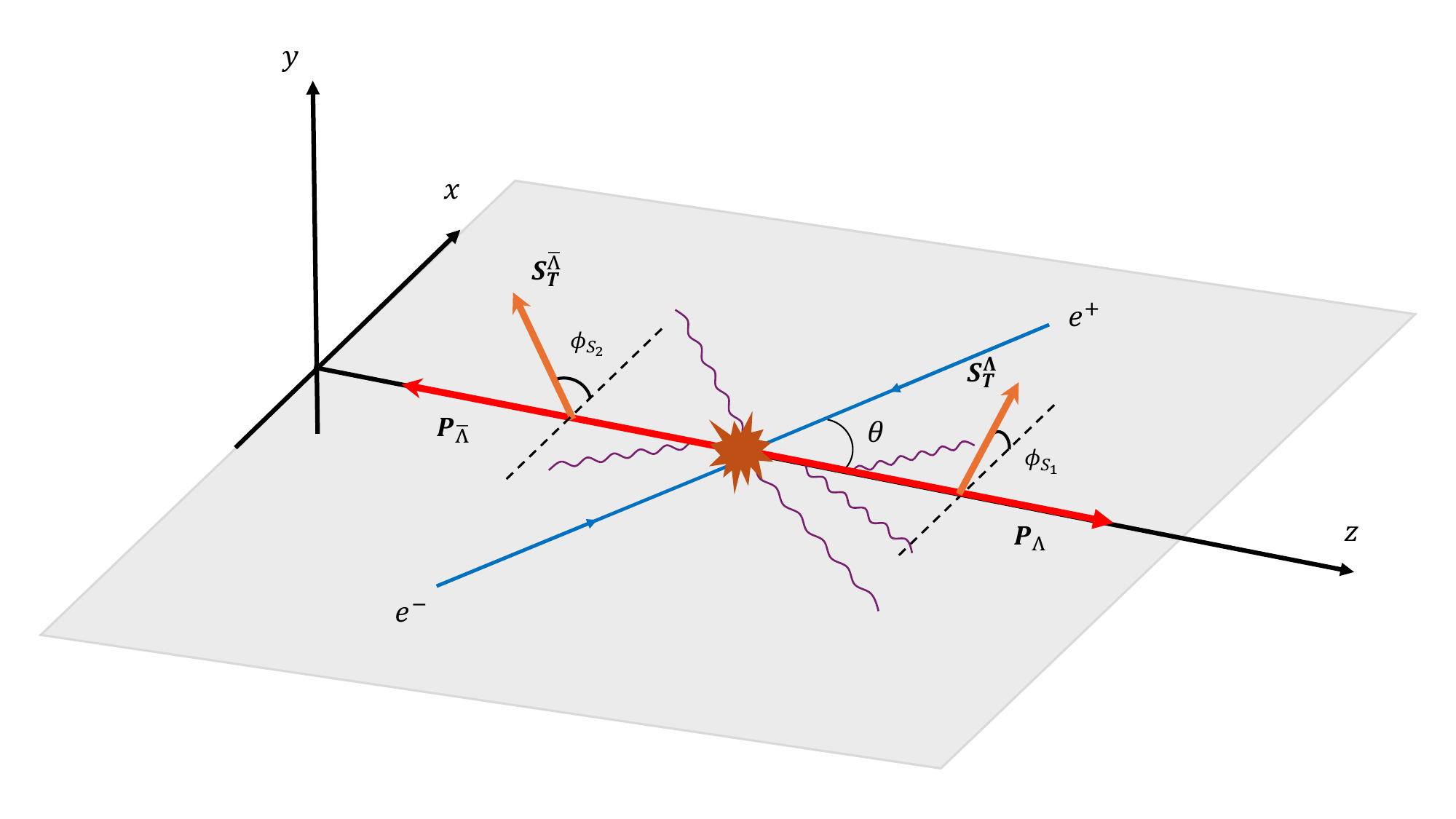}
    \caption{Kinematic configuration for $e^+e^- \to \Lambda\bar{\Lambda} X$. The transverse spin vectors $\boldsymbol{S}_T^{\Lambda}$ and $\boldsymbol{S}_T^{\bar{\Lambda}}$ have azimuthal angles $\phi_{S_1}$ and $\phi_{S_2}$ in the lab frame. The electron-positron collision axis is oriented at an angle $\theta$ relative to the $z$-axis.}
    \label{fig:kinematics}
\end{figure}

We begin with the parton model description of inclusive hyperon pair production in electron-positron annihilation
\begin{align}\label{eq:pro}
    e^{-}(\ell) + e^{+}(\ell') \rightarrow \Lambda ({P}^{ \Lambda},\boldsymbol{S}^{\Lambda}) + \bar\Lambda ({P}^{\bar \Lambda},\boldsymbol{S}^{\bar\Lambda}) + X\,,
\end{align}
where $\ell$ and $\ell'$ are the four-momenta of the incoming $e^-$ and $e^+$, respectively. The final state consists of a $\Lambda$ hyperon with four momentum $P^\Lambda$ and spin three vector $\boldsymbol{S}^{\Lambda}$, an $\bar\Lambda$ with corresponding variables ${P}^{\bar \Lambda}$ and $\boldsymbol{S}^{\bar\Lambda}$, and any additional undetected particles, which are denoted by $X$. In the parton model calculation, the hyperons are produced via the fragmentation of the primary quark and anti-quark pairs. 

This section begins by establishing the relevant kinematics, followed by the definition of the cross section of polarized hyperon pair production in terms of polarized FFs, and concludes by re-expressing this result using the fragmenting density matrices (FDMs).

\subsection{Kinematics}\label{subsec:kinematics}

We define the lab frame such that $\Lambda$ hyperon moves along the $+\hat{z}$ direction and $\bar \Lambda$ along the $-\hat{z}$ direction. The incoming electron-positron axis lies in the $xz$-plane, as depicted in figure~\ref{fig:kinematics}. The total momentum transfer is $q^\mu = \ell^\mu + \ell'^{\mu}$, with invariant mass $Q \equiv \sqrt{q^2}$. 

While kinematics are defined in the lab frame, it is more convenient to parameterize the hyperon spin vectors $\boldsymbol{S}^{\Lambda}$ and $\boldsymbol{S}^{\bar{\Lambda}}$ in their respective helicity frames, where the quantization axis is aligned with each particle's momentum direction. For $\Lambda$, its helicity frame axes coincide with those of the lab frame $\left(\hat{x}, \hat{y}, \hat{z}\right)$. For $\bar \Lambda$, the helicity frame is defined by $\left(\hat{\bar{x}}, \hat{\bar{y}}, \hat{\bar{z}}\right) = \left(\hat{x}, -\hat{y}, -\hat{z}\right)$ \cite{Chen:1994ar, Jacob:1959at, DAlesio:2021dcx, Anselmino:2005sh, Batozskaya:2023rek, Perotti:2018wxm}. In these helicity frames, the normalized ($|\boldsymbol{S}|^2=1$) spin vectors are parameterized by their helicities $\lambda_{1,2}$ and azimuthal angles $\varphi_{1,2}$ as
\begin{align}
\boldsymbol{S}^\Lambda = \left(|\boldsymbol{S}_T^{\Lambda}|\,\cos\varphi_{1},|\boldsymbol{S}_T^{\Lambda}|\,\sin\varphi_{1},\lambda_1\right), \quad \boldsymbol{S}^{\bar\Lambda}=\left(|\boldsymbol{S}_T^{\bar\Lambda}|\,\cos\varphi_{2},|\boldsymbol{S}_T^{\bar\Lambda}|\,\sin\varphi_{2},\lambda_2\right)\,.
\end{align}
where the helicity-frame angles are related to the observable lab-frame angles $\phi_{S_{1,2}}$ (defined in figure~\ref{fig:kinematics}) by $\varphi_1=\phi_{S_1}$ and $\varphi_2=2\pi-\phi_{S_2}$~\cite{Anselmino:2005sh}. Accordingly, the projection of the spin vector $\boldsymbol{S}^{\Lambda}$ onto the lab-frame $z$-axis is its helicity, $P^{\Lambda}_z = \lambda_1$. For the $\bar{\Lambda}$, which moves in the opposite direction, the projection is $P^{\bar \Lambda}_z = -\lambda_2$.

\subsection{Polarized fragmentation functions}\label{subsec:polarized_ff}

Within the parton model, the differential cross section for producing the polarized hyperon pair can be decomposed into unpolarized ($U$), longitudinal ($L$), and transverse ($T$) components as follows~\cite{Boer:1997mf}
\begin{align}\label{eq:parton_model_pol_sec}
\frac{\mathrm{d} \sigma(\boldsymbol{S}^{\Lambda},\boldsymbol{S}^{\bar\Lambda})}{\mathrm{d} z_1 \, \mathrm{d} z_2\,\mathrm{d} \Omega} = 
\sum_q e_q^2 
\Biggl[ &
  \frac{\mathrm{d} \sigma_0^{U}}{\mathrm{d} \Omega} \, \mathcal{D}^{U}_{\Lambda/q}(z_1,\mu)\, \mathcal{D}^{U}_{\bar{\Lambda}/\bar{q}}(z_2,\mu)\,
 +P_z^{\Lambda} P_z^{\bar\Lambda}\,\frac{\mathrm{d} \sigma_0^{L}}{\mathrm{d} \Omega} \, \mathcal{D}^{L}_{\Lambda/q}(z_1,\mu)\, \mathcal{D}^{L}_{\bar{\Lambda}/\bar{q}}(z_2,\mu)\,
\nonumber\\
+ & |\boldsymbol{S}_T^\Lambda||\boldsymbol{S}_T^{\bar\Lambda}|\cos( \phi_{S_1} + \phi_{S_2})\, \frac{\mathrm{d} \sigma_0^{T}}{\mathrm{d} \Omega} \, \mathcal{D}^{T}_{\Lambda/q}(z_1,\mu)\, \mathcal{D}^{T}_{\bar{\Lambda}/\bar{q}}(z_2,\mu)\,
\Biggr]\,,
\end{align}
where the sum $\sum_q$ runs over all light quark flavors, and $e_q$ denotes the fractional charge of quarks. $\mathcal{D}^{\mathcal{P}}_{\Lambda/q}(z, \mu)$ is the FF describing the fragmentation of a quark $q$ into a $\Lambda$ hyperon. The momentum fractions carried by the $\Lambda$ and $\bar{\Lambda}$ are defined as $z_1 \equiv 2P^{\Lambda}\cdot q/Q^2$ and $z_2 \equiv 2P^{\bar{\Lambda}}\cdot q/Q^2$, respectively.

At leading order, $e^+e^-\to\gamma^* \to q\bar q$, the partonic cross sections for different polarization modes are given by
\begin{align}\label{eq:LO_xsec}
     \frac{\mathrm{d} \sigma_0^U}{\mathrm{d}\Omega} = \frac{\mathrm{d} \sigma_0^L}{\mathrm{d}\Omega} = \frac{N_c\,\alpha_{em}^2}{4Q^2}(1+\cos^2\theta)\,, \quad\quad
     \frac{\mathrm{d} \sigma_0^T}{\mathrm{d}\Omega} = \frac{N_c\,\alpha_{em}^2}{4Q^2}\sin^2\theta\,,
\end{align}
where $\theta$ is the scattering angle defined in the figure~\ref{fig:kinematics}, $N_c=3$ is the number of color, and $\alpha_{em}$ is the fine structure constant.

The FFs, $\mathcal{D}^{\mathcal{P}}_{\Lambda/q}(z, \mu)$, describe the number density of $\Lambda$ hyperons produced from the fragmentation of a quark of flavor $q$, carrying a momentum fraction $z$. The superscript $\mathcal{P} \in \{U, L, T\}$ specifies the spin structure of the process, corresponding to unpolarized, longitudinally polarized, or transversely polarized quarks and the resulting hadron polarization. At the operator level, they are defined as light-cone correlators of gauge-invariant collinear quark fields, $\chi_n$~\cite{Barone:2003fy, Jain:2011xz, Chen:1994ar}\footnote{Our notation $\{\mathcal{D}^U, \mathcal{D}^L, \mathcal{D}^T\}$ is related to other common conventions in the literature. It corresponds to the Jaffe-Ji convention $\{f_1, g_1, h_1\}$~\cite{Jaffe:1991ra,Chen:1994ar} and the notation $\{D, \Delta D, \Delta_T D\}$ used in Refs.~\cite{Barone:2003fy,Kang:2023elg,Collins:1992kk}.}:
\begin{align} \label{eq:FFs}
\mathcal{D}^U_{\Lambda/q}(z) &= \frac{1}{z} \int \frac{\mathrm{d} y_+}{4\pi} \, e^{ i k_- y_+ / 2} \sum_X \frac{1}{2N_c} \text{Tr}\left[\frac{\slashed{\bar{n}}}{2} \langle 0 | \chi_n(y_+)| \Lambda X \rangle \langle \Lambda X | \bar{\chi}_n(0) | 0 \rangle \right]\,,  \\
\mathcal{D}^L_{\Lambda/q}(z) &= \frac{1}{z} \int \frac{\mathrm{d} y_+}{4\pi} \, e^{ i k_- y_+ / 2} \sum_X \frac{1}{2N_c} \text{Tr}\left[\frac{\slashed{\bar{n}}}{2} \gamma_5 \langle 0 | \chi_n(y_+)| \Lambda X \rangle \langle \Lambda X | \bar{\chi}_n(0) | 0 \rangle \right]\,, \nn \\
{S}^i_{\perp} \mathcal{D}^T_{\Lambda/q}(z) &= \frac{1}{z} \int \frac{\mathrm{d} y_+}{4\pi} \, e^{ i k_- y_+ / 2} \sum_X \frac{1}{2N_c} \text{Tr}\left[\frac{\slashed{\bar{n}}}{2} \gamma_\perp^i \gamma_5 \langle 0 | \chi_n(y_+)| \Lambda X \rangle \langle \Lambda X | \bar{\chi}_n(0) | 0 \rangle \right]\,.\notag
\end{align}
Here, the integral is over the light-cone coordinate $y_+$, and the momentum fraction is defined as $z = P^\Lambda_-/k_-$, where $k_-$ is the large light-cone component of the fragmenting quark's momentum. The Dirac structures involving $\gamma_5$ and $\gamma_\perp^i$ project out the longitudinal and transverse spin components, respectively. For the transverse component, the spin four-vector is given by ${S}^i_{\perp}=\left(0,\boldsymbol{S}_T^\Lambda,0\right)$. The sum is performed over all unobserved final-state remnants, collectively denoted by $X$.

The above definitions utilize light-cone coordinates, in which a four-vector $p^\mu$ is decomposed as
\begin{align}
    p^\mu = p_+ \frac{\bar{n}^\mu}{2} + p_- \frac{n^\mu}{2} + p_\perp^\mu\,,
\end{align}
using the light-like basis vectors $n^\mu = (1, 0, 0, 1)$ and $\bar{n}^\mu = (1, 0, 0, -1)$~\cite{Becher:2014oda}. The gauge-invariant collinear quark field, $\chi_n$, is constructed from the collinear component of the full quark field
\begin{align}
    \chi_n = W_n^\dagger \xi_n\,, \qquad \text{where} \qquad \xi_n = \frac{\slashed{n} \slashed{\bar{n}}}{4} \psi\,.
\end{align}
Gauge invariance is ensured by the collinear Wilson line, $W_n$, which is a path-ordered exponential of the collinear gluon field
\begin{align}
    W_n(x) = \mathbf{P} \exp\left[i g_s \int_{-\infty}^0 \mathrm{d} s\, \bar{n} \cdot A_c(x + s \bar{n}) \right]\,.
\end{align}

\subsection{Fragmenting density matrices}\label{subsec:FDM}

In this section, we rederive Eq.~\eqref{eq:parton_model_pol_sec} using the helicity formalism. This method provides a clear probabilistic interpretation of spin-dependent processes and serves as a powerful alternative to standard techniques~\cite{Boer:1997mf}. In this approach, the central object is the production density matrix, which is systematically constructed from the helicity amplitudes of the final state~\cite{DAlesio:2021dcx}.

At Born level, the dynamics of the process $e^+e^- \to q\bar{q}$ are encapsulated in an abstract amplitude vector, $|\mathcal{M}\rangle$, which lives in the spin space of the final-state $q\bar{q}$ pair. The helicity amplitude for a specific final-state helicity configuration, $ \{\lambda_q, \lambda_{\bar{q}}\}$, is the projection of this vector onto the corresponding basis state
\begin{align}
  \mathcal{M}_{\lambda_q, \lambda_{\bar{q}}} \equiv \langle\lambda_q, \lambda_{\bar{q}} | \mathcal{M}\rangle\,.
\end{align}
The unpolarized cross section is proportional to the squared norm of this vector, $\langle\mathcal{M}|\mathcal{M}\rangle$. However, the full spin correlation structure is contained in the production density matrix. Its components in the helicity basis are constructed from the outer product of the amplitude vector with itself
\begin{align} \label{eq:production_density_matrix_LO}
\left(\frac{\md \hat \sigma_0}{\md \Omega}\right)_{\lambda_{q}\lambda_{q}^{\prime},\lambda_{\bar{q}}\lambda^{\prime}_{\bar{q}} }
&= \sum_{q}\frac{N_c}{16\pi^2 Q^2} \langle \lambda_q, \lambda_{\bar q}| \mathcal{M}\rangle \langle \mathcal{M}| \lambda_q^\prime, \lambda_{\bar q}^\prime\rangle \notag \\
& = \sum_{q}\frac{N_c}{16\pi^2 Q^2}
\mathcal{M}_{\lambda_{q}\lambda_{\bar{q}}}
\mathcal{M}^*_{\lambda^{\prime}_{q}\lambda^{\prime}_{\bar{q}}}\,.
\end{align}
This production density matrix can be conveniently decomposed onto a basis of spin operators \cite{Brandenburg:1996df, Afik:2022kwm, Jaffe:1996zw}
\begin{align} \label{eq:density_matrix_partonic}
\left(\frac{\md \hat\sigma_0}{ \md \Omega}\right)_{\lambda_q\lambda_q',\lambda_{\bar{q}}\lambda_{\bar{q}}'}
= \sum_{q}{e_q^2}&\left[  \frac{\mathrm{d} \sigma_0^U}{\mathrm{d}\Omega}\,
\hat{I}^{q}_{\lambda_q\lambda_q'} \otimes \hat{I}^{\bar{q}}_{\lambda_{\bar{q}}\lambda_{\bar{q}}'}  - 
\frac{\mathrm{d} \sigma_0^L}{\mathrm{d}\Omega}
(\hat{\sigma}^{q}_z)_{\lambda_q\lambda_q'} \otimes (\hat{\sigma}^{\bar{q}}_{\bar{z}})_{\lambda_{\bar{q}}\lambda_{\bar{q}}'}\right. \nonumber \\
&\left.  + \frac{\mathrm{d} \sigma_0^T}{\mathrm{d}\Omega} 
 (\hat{\sigma}^{q}_x)_{\lambda_q\lambda_q'} \otimes (\hat{\sigma}^{\bar{q}}_{\bar{x}})_{\lambda_{\bar{q}}\lambda_{\bar{q}}'}+ \frac{\mathrm{d} \sigma_0^T}{\mathrm{d}\Omega} 
 (\hat{\sigma}^{q}_y)_{\lambda_q\lambda_q'} \otimes (\hat{\sigma}^{\bar{q}}_{\bar{y}})_{\lambda_{\bar{q}}\lambda_{\bar{q}}'}\right]\,.
\end{align}
Here, $\hat{I}^{q,\bar q}$ is the $2\times2$ identity matrix, $\hat{\sigma}^q_i$ and $\hat{\sigma}^{\bar{q}}_{\bar{i}}$ denote the spin operators projected along the $i$-th and $\bar{i}$-th axes in the helicity frames of the quark $q$ and anti-quark $\bar{q}$, respectively. Throughout this work, we use the superscript to indicate the particle, and the subscript ($i = x, y, z$ or $\bar{i} = \bar{x}, \bar{y}, \bar{z}$) to label the spatial direction in its corresponding helicity frame.

To obtain the production density matrix for the final-state hyperons, the partonic production process must be dressed with the subsequent fragmentation of the quarks. While a spin-averaged calculation in the parton model involves a convolution with standard FFs, a full spin analysis requires the use of FDMs~\cite{DAlesio:2021dcx, DAlesio:2022wbc, Anedda:2025bts}.

The FDM describes the spin transfer during the fragmentation process. Schematically, it is constructed from the outer product of the collinear amplitude, $\mathcal{M}_c$, for a quark fragmenting into a hyperon, $q \to \Lambda + X$:
\begin{align}
    \hat D_{\lambda_q^\prime \lambda_q, \lambda_\Lambda \lambda_\Lambda^\prime} \propto \langle \lambda_q^\prime, \lambda_\Lambda| \mathcal{M}_c \rangle \langle \mathcal{M}_c|\lambda_q, \lambda_\Lambda^\prime\rangle \,.
\end{align}
The full production density matrix for the $\Lambda\bar{\Lambda}$ pair is then constructed by contracting the partonic production density matrix from Eq.~\eqref{eq:production_density_matrix_LO} with the FDMs for the quark and antiquark fragmentation channels. The sum is performed over all repeated partonic helicity indices
\begin{align} \label{eq:hadronic_density_matrix}
\left(\frac{\md \hat \sigma}{\md z_1 \, \md z_2\, \md \Omega}\right)_{\lambda_{\Lambda}\lambda_{\Lambda}^{\prime},\lambda_{\bar{\Lambda}}\lambda^{\prime}_{\bar{\Lambda}} }
= \sum_{q}\sum_{\{\lambda, \lambda'\}} & \frac{N_c}{16\pi^2 Q^2}
\mathcal{M}_{\lambda_{q}\lambda_{\bar{q}}}
\mathcal{M}^*_{\lambda^{\prime}_{q}\lambda^{\prime}_{\bar{q}}} \nonumber \\
& \times \hat{D}_{\lambda^{\prime}_{q}\lambda_{q},\lambda_{\Lambda}\lambda^{\prime}_{\Lambda}}(z_1,\mu)
\hat{D}_{\lambda^{\prime}_{\bar{q}}\lambda_{\bar{q}},\lambda_{\bar{\Lambda}}\lambda^{\prime}_{\bar{\Lambda}}}(z_2,\mu)\,.
\end{align}

The FDM, $\hat{D}_{\lambda'_q\lambda_q,\lambda_\Lambda\lambda'_\Lambda}$, can be decomposed onto the same operator basis used for the partonic density matrix in Eq.~\eqref{eq:density_matrix_partonic}. In this basis, the expansion coefficients are the polarized FFs~\cite{Chen:1994ar, Jaffe:1996wp, Jaffe:1996zw}
\begin{align}
\hat{D}_{\lambda_q^\prime\lambda_{q},\lambda_{\Lambda}\lambda_{\Lambda}^\prime}(z)
&= \frac{1}{2}\left\{\mathcal{D}^U_{\Lambda/q}(z)\, \hat{I}^q_{\lambda_q^\prime\lambda_{q}} \otimes \hat{I}^{\Lambda}_{\lambda_\Lambda\lambda_{\Lambda}^\prime}
+ \mathcal{D}^L_{\Lambda/q}(z)\, (\hat{\sigma}^{q}_z)_{\lambda_q^\prime\lambda_{q}} \otimes (\hat{\sigma}^{\Lambda}_z)_{\lambda_\Lambda\lambda_{\Lambda}^\prime} \right.\nonumber \\
&\left.\quad + \mathcal{D}^T_{\Lambda/q}(z) \left[
(\hat{\sigma}^{q}_x)_{\lambda_q^\prime\lambda_{q}} \otimes (\hat{\sigma}^{\Lambda}_x)_{\lambda_\Lambda\lambda_{\Lambda}^\prime}
+ (\hat{\sigma}^{q}_y)_{\lambda_q^\prime\lambda_{q}} \otimes (\hat{\sigma}^{\Lambda}_y)_{\lambda_\Lambda\lambda_{\Lambda}^\prime}
\right]\right\}\,.
\end{align}
Substituting this decomposition into the expression for the production density matrix, Eq.~\eqref{eq:hadronic_density_matrix}, and summing over the intermediate quark helicities yields the final expression for the observable $\Lambda\bar{\Lambda}$ system
\begin{align} \label{eq:density_matrix}
\left(\frac{\md \hat\sigma}{ \md z_1 \, \md z_2\, \md \Omega}\right)_{\lambda_\Lambda\lambda_\Lambda',\lambda_{\bar{\Lambda}}\lambda_{\bar{\Lambda}}'}
&= \frac{\md \sigma^U}{ \md z_1 \, \md z_2\, \md \Omega}\,
\hat{I}^{\Lambda}_{\lambda_\Lambda\lambda_\Lambda'} \otimes \hat{I}^{\bar\Lambda}_{\lambda_{\bar{\Lambda}}\lambda_{\bar{\Lambda}}'}  - \frac{\md \sigma^L}{ \md z_1 \, \md z_2\, \md \Omega}\,
(\hat{\sigma}^{\Lambda}_z)_{\lambda_\Lambda\lambda_\Lambda'} \otimes (\hat{\sigma}^{\bar\Lambda}_{\bar{z}})_{\lambda_{\bar{\Lambda}}\lambda_{\bar{\Lambda}}'}\nonumber \\
&\hspace{-5.5em}+ \frac{\md \sigma^T}{ \md z_1 \, \md z_2\, \md \Omega}
(\hat{\sigma}^{\Lambda}_x)_{\lambda_\Lambda\lambda_\Lambda'} \otimes (\hat{\sigma}^{\bar\Lambda}_{\bar{x}})_{\lambda_{\bar{\Lambda}}\lambda_{\bar{\Lambda}}'}
+ \frac{\md \sigma^T}{ \md z_1 \, \md z_2\, \md \Omega}
(\hat{\sigma}^{\Lambda}_y)_{\lambda_\Lambda\lambda_\Lambda'} \otimes (\hat{\sigma}^{\bar\Lambda}_{\bar{y}})_{\lambda_{\bar{\Lambda}}\lambda_{\bar{\Lambda}}'}\,,
\end{align}
where we define the spin-dependent contribution to the cross section as
\begin{align} \label{eq:LO_pol_cross_section}
    \frac{\mathrm{d} \sigma^{\mathcal{P}}}{\mathrm{d} z_1 \, \mathrm{d} z_2\,\mathrm{d} \Omega} = \sum_q e_q^2 \frac{\mathrm{d} \sigma_0^{\mathcal{P}}}{\mathrm{d} \Omega} \, \mathcal{D}^{\mathcal{P}}_{\Lambda/q}(z_1,\mu)\, \mathcal{D}^{\mathcal{P}}_{\bar{\Lambda}/\bar{q}}(z_2,\mu)\,.
\end{align}
To obtain a cross section differential in the spin vectors $\boldsymbol{S}^{\Lambda}$ and $\boldsymbol{S}^{\bar{\Lambda}}$, we contract the production density matrix from Eq.~\eqref{eq:density_matrix} with the helicity density matrices for the individual polarized hyperons. The helicity density matrix for each hadron has a standard decomposition in terms of its spin vector $\boldsymbol{S}$ 
\begin{align} \label{eq:hadron_helicity_density_matrix}
\hat \rho^{\boldsymbol{S}^{\Lambda}}_{\lambda_{\Lambda}\lambda_{\Lambda}^{\prime}}
&= \frac{1}{2}\left(\hat{I}^{\Lambda}_{\lambda_\Lambda\lambda_{\Lambda}^\prime} + \boldsymbol{S}^{\Lambda} \cdot \hat{\boldsymbol{\sigma}}^{\Lambda}_{\lambda_\Lambda\lambda_{\Lambda}^\prime}\right),\nonumber \\
\hat \rho^{\boldsymbol{S}^{\bar{\Lambda}}}_{\lambda_{\bar{\Lambda}}\lambda_{\bar{\Lambda}}^{\prime}}
&= \frac{1}{2}\left(\hat{I}^{\bar\Lambda}_{\lambda_{\bar{\Lambda}}\lambda_{\bar{\Lambda}}^\prime} + \boldsymbol{S}^{\bar{\Lambda}} \cdot \hat{\boldsymbol{\sigma}}^{\bar{\Lambda}}_{\lambda_{\bar{\Lambda}}\lambda_{\bar{\Lambda}}^\prime}\right)\,,
\end{align}
where $\hat{\boldsymbol{\sigma}}^{\Lambda} = (\hat{\sigma}^{\Lambda}_x, \hat{\sigma}^{\Lambda}_y, \hat{\sigma}^{\Lambda}_z)$ and $\hat{\boldsymbol{\sigma}}^{\bar{\Lambda}} = (\hat{\sigma}^{\bar{\Lambda}}_{\bar{x}}, \hat{\sigma}^{\bar{\Lambda}}_{\bar{y}}, \hat{\sigma}^{\bar{\Lambda}}_{\bar{z}})$. The hadron helicity density matrix describes the spin orientation of the particle in its helicity frame, with the relation $\operatorname{Tr}(\hat \sigma_i \hat \rho)=S_i$ gives the $i$-th component of the spin vector $\boldsymbol{S}$ in the helicity frame of the particle \cite{DAlesio:2021dcx, Anselmino:2005sh}.

Therefore, contracting the production density matrix with the hadron helicity density matrices projects it onto a definite spin configuration specified by the spin vectors $\boldsymbol{S}^{\Lambda}$ and $\boldsymbol{S}^{\bar{\Lambda}}$ defined in their respective helicity frames. The contraction is performed as follows
\begin{align} \label{eq:polarized_cross_section}
\md\sigma(\boldsymbol{S}^{\Lambda},\boldsymbol{S}^{\bar\Lambda})
= \text{Tr}\left[ \hat \rho^{\boldsymbol{S}^{\Lambda}} \otimes \hat \rho^{\boldsymbol{S}^{\bar{\Lambda}}} \, \md \hat \sigma\right]\,.
\end{align}
Performing the trace explicitly using the decompositions in Eq.~\eqref{eq:density_matrix} and Eq.~\eqref{eq:hadron_helicity_density_matrix} recovers the polarized cross section previously given in Eq.~\eqref{eq:parton_model_pol_sec}, demonstrating the self-consistency of the formalism.

\subsection{Spin correlations}

With the cross section decomposed into longitudinal and transverse spin components, we can define spin correlations that isolate these contributions. These correlations are key observables for probing the spin structure of the fragmentation process.

The longitudinal correlation, $C_{LL}$, probes correlations between the helicities of the produced hyperons. It is defined by comparing the production rates where the $\Lambda$ and $\bar{\Lambda}$ have aligned versus anti-aligned longitudinal spins. A non-zero $C_{LL}$ therefore indicates a sensitivity to the helicity-dependent dynamics of the underlying partonic interaction and fragmentation.

In contrast, the transverse correlation, $C_{TT}$, is sensitive to spin correlations in the plane transverse to the hyperons' motion. It compares production rates for configurations with different relative orientations of the transverse spin vectors. Since transverse spin phenomena often arise from spin-orbit couplings and quantum interference, $C_{TT}$ provides valuable insight into the more intricate aspects of QCD spin dynamics during hadronization.

While the longitudinal spin correlation, $C_{LL}$, is straightforward to define, the transverse spin correlation, $C_{TT}$, depends on the choice of quantization axes used to measure the transverse spins, appearing in the form of $\cos(\phi_{S_1}+\phi_{S_2})$ in Eq.~\eqref{eq:parton_model_pol_sec}. As a result, the transverse spin correlation reaches its maximum when both quantization axes lie either within the event plane (with $C_{TT} > 0$) or perpendicular to the event plane (with $C_{TT}< 0$). To avoid complications in theoretical analysis, we define $C_{TT}$ in the following way to eliminate the azimuthal angle dependence. In experimental measurements, one can directly access $C_{TT}$ by measuring along these two specific directions. The longitudinal $C_{LL}$ and transverse $C_{TT}$ spin correlations are defined as 
\begin{align} \label{eq:asym}
C_{LL} &\equiv \frac{\md \sigma(P_z^{\Lambda}= \pm 1,P_z^{\bar\Lambda}=\pm 1) - \md \sigma(P_z^{\Lambda}=\pm 1,P_z^{\bar\Lambda}=\mp 1)}{\md \sigma(P_z^{\Lambda}=\pm 1,P_z^{\bar\Lambda}=\pm 1) + \md \sigma(P_z^{\Lambda}= \pm 1,P_z^{\bar\Lambda}=\mp 1)} = \frac{\md \sigma^L}{\md \sigma^U}\,, \\
C_{TT} &\equiv \frac{2\int_0^{2\pi}\md\phi_{S_1} \int_0^{2\pi}\md\phi_{S_2} \cos{(\phi_{S_1} + \phi_{S_2})}  \, \md \sigma(\boldsymbol{S}_T^\Lambda,\boldsymbol{S}_T^{\bar{\Lambda}}) }{\int_0^{2\pi}\md\phi_{S_1} \int_0^{2\pi}\md\phi_{S_2} \, \md \sigma(\boldsymbol{S}_T^\Lambda,\boldsymbol{S}_T^{\bar{\Lambda}}) } =\frac{\md \sigma^T}{\md \sigma^U}\,. 
\end{align}
Within the parton model, where fragmentation is described by FFs, these correlations can be computed explicitly. Substituting the expressions from Eq.~\eqref{eq:LO_pol_cross_section} into the definitions above yields
\begin{align} \label{eq:asym-FF}
C_{LL} &= \frac{\sum_q e_q^2\, \mathcal{D}^L_{\Lambda/q}(z_1)\, \mathcal{D}^L_{\bar{\Lambda}/\bar{q}}(z_2)}{
\sum_q e_q^2\, \mathcal{D}^U_{\Lambda/q}(z_1)\, \mathcal{D}^U_{\bar{\Lambda}/\bar{q}}(z_2)}\,, \\
C_{TT} &= \frac{\sum_q e_q^2\, \sin^2\theta\, \mathcal{D}^T_{\Lambda/q}(z_1)\, \mathcal{D}^T_{\bar{\Lambda}/\bar{q}}(z_2)}{
\sum_q e_q^2\, (1+\cos^2\theta)\, \mathcal{D}^U_{\Lambda/q}(z_1)\, \mathcal{D}^U_{\bar{\Lambda}/\bar{q}}(z_2)}\,.
\end{align}

\section{Factorization and resummation with a thrust cut}\label{sec:thrust}

To provide precise theoretical predictions for spin correlations in the dijet limit, we go beyond the fixed-order parton model by developing a factorization theorem within SCET. This allows for the systematic resummation of large logarithms of the thrust variable $T$.

Thrust, $T$, is a classic event shape observable at lepton colliders \cite{Farhi:1977sg}, defined by
\begin{align}
T \equiv \max _{\hat{n}} \frac{\sum_i\left|\hat{n} \cdot \vec{p}_i\right|}{\sum_i\left|\vec{p}_i\right|}\,,
\end{align}
where the sum runs over the momenta $\vec p_i$ of all final-state particles, and the unit vector $\hat n$ that maximizes this expression defines the thrust axis. The variable $\tau \equiv 1-T$ is used to characterize the event topology. The limit $\tau \to 0$ corresponds to the pencil-like dijet configurations, whereas the opposite limit, $\tau \to 1/2$, describes spherical, isotropic events. 

A plane perpendicular to the thrust axis divides the event into two hemispheres, denoted $\mathcal{H}_{L}$ and $\mathcal{H}_{R}$. In the dijet limit ($\tau \ll 1$), the thrust variable can be decomposed as
\begin{align}
\tau=\frac{p_L^2+p_R^2}{Q^2}+\frac{k}{Q}\,,
\end{align}
where $p_L^2$ and $p_R^2$ represent the invariant masses of the collinear radiations in the left and right hemispheres,  and $k$ is the contribution from soft radiation in both hemispheres.

The factorization of the thrust distribution in the dijet limit into hard, jet, and soft functions is well established within the framework of SCET~\cite{Becher:2008cf, Schwartz:2014sze}. To describe the production of an identified hadron within a jet, this framework requires that the standard jet function be replaced by a FJF. The semi-inclusive and exclusive FJFs were developed in Ref.~\cite{Kang:2016ehg, Kang:2016mcy, Jain:2011iu, Kang:2023elg} with both unpolarized and polarized fragmenting hadrons using the clustering jet algorithm. The formalism for unpolarized FJFs with a thrust cut was proposed in Ref.~\cite{Procura:2009vm, Jain:2011xz}. In this work, we generalize this framework with a thrust cut to describe the production of polarized hadrons.

\subsection{Factorization with a thrust cut}

The required spin dependence for the polarized cross section resides entirely within the polarized FJFs, as both the hard function and the thrust soft function are spin-independent. The resulting factorization theorem takes the form \cite{Catani:1992ua, Korchemsky:1999kt, Fleming:2007qr, Schwartz:2007ib}
\begin{align}\label{eq:fac}
\frac{\md \sigma^\mathcal{P}}{\md \tau \,  \md z_1 \, \md z_2\, \md \Omega}
=\frac{\md \sigma^\mathcal{P}_0}{\md\Omega} H(Q^2,\mu)&\int {\md p^2_L}\,{\md p^2_R}\, \md k \,\delta\left(\tau-\frac{p^2_L+p^2_R}{Q^2}-\frac{k}{Q}\right)\nn\\
&\times S_T\left(k,\mu\right) \sum_{q} {{e_q^2}} \, \mathcal{G}^\mathcal{P}_{\Lambda/q}\left(z_1,p^2_L,\mu\right) \mathcal{G}^\mathcal{P}_{\bar\Lambda/\bar{q}}\left(z_2,p^2_R,\mu\right)\,,
\end{align}
where the Born differential cross section $\md \sigma^\mathcal{P}_0/\md\Omega$ is defined in Eq.~\eqref{eq:LO_xsec}. $H(Q^2,\mu)$ denotes the hard function, containing high-energy virtual corrections at the scale $Q$. The thrust soft function, $S_T(k,\mu)$, describes wide-angle soft emissions. The functions $\mathcal{G}^\mathcal{P}_{\Lambda/q}$ and $\mathcal{G}^\mathcal{P}_{\bar{\Lambda}/\bar{q}}$ are the polarized FJFs, where the superscript $\mathcal{P} \in \{ U, L, T\}$ specifies the unpolarized, longitudinal, or transverse polarization states of the hyperon, respectively. All functions depend on the standard factorization scale 
$\mu$, which governs their renormalization group evolution.

The convolution structure of Eq.~\eqref{eq:fac}, arising from the delta function, complicates the all-order resummation of large logarithms in momentum space. This difficulty can be circumvented by performing a Laplace transform~\cite{Becher:2006nr},
\begin{align}\label{eq:laptransf}
    {F}(u) = \int_0^\infty \md \tau \, e^{-{u\tau}/{e^{\gamma_E}}} F(\tau),\quad\quad {F}(\tau) =\int_{\gamma-i\infty}^{\gamma+i\infty} \frac{\md u}{2\pi i}\,\frac{ e^{{u\tau}/{e^{\gamma_E}}}}{e^{\gamma_E}}  F(u)\,,
\end{align}
under which the convolution becomes a simple product. The factorized cross section is then expressed as
\begin{align}\label{eq:Laplace_fac}
\frac{\md \sigma^{\mathcal{P}}}{\md \tau \,  \md z_1 \, \md z_2\, \md \Omega}
=\frac{\md \sigma^{\mathcal{P}}_0}{\md\Omega}  
H\left(Q^2,\mu\right) &\int_{\gamma- i\infty}^{\gamma+ i\infty}\frac{\md u}{2\pi i}\frac{e^{u\tau/{e^{\gamma_E}}}}{e^{\gamma_E}}{S}_T\left(\frac uQ,\mu\right)\,\nn\\
&\times\sum_{q} {e_q^2}\,{\mathcal{G}}^{\mathcal{P}}_{{\Lambda}/{q} }\left(z_1,\frac u{Q^2},\mu\right) {\mathcal{G}}^{\mathcal{P}}_{\bar{\Lambda}/\bar q}\left(z_2,\frac u{Q^2},\mu\right)\,.
\end{align}
The hard and soft functions, along with their required resummation ingredients, are identical to those for the standard inclusive thrust distribution. For completeness, their expressions needed for resummation at NNLL accuracy are summarized in appendix~\ref{app:func}. The new ingredient of this work is the polarized FJF. As this is the first calculation of this function, we dedicate the following subsection to its explicit calculation.

\subsection{Fragmenting jet functions}\label{sec:FJFs}

The FJFs describe the fragmentation of a parton into a hadron, differential in both the hadron's momentum fraction, $z$, and the jet's invariant mass, $s$. At the operator level, they are defined via the correlation function of gauge-invariant collinear fields~\cite{Jain:2011xz, Kang:2023elg, Procura:2009vm}
\begin{align}\label{eq:FJFs}
&\mathcal{G}^U_{\Lambda/q}(z,s)=\frac{1}{P^{\Lambda}_-}\int\frac{\md y_-}{4\pi} \, e^{ ik_+y_-/2} \sum_X\frac{1}{2N_c} \mathrm{tr}\left[\frac{\slashed{\bar{n}}}{2}\langle0|[\delta(\omega-\overline{\mathcal{P}})\chi_n(y_-)]|\Lambda X\rangle\langle \Lambda X|\bar{\chi}_n(0)|0\rangle\right]\,,\notag\\
&\mathcal{G}^L_{\Lambda/q}(z,s)=\frac{1}{P^{\Lambda}_-}\int\frac{\md y_-}{4\pi} \, e^{ ik_+y_-/2}  \sum_X \frac{1}{2N_c} \mathrm{tr}\left[\frac{\slashed{\bar{n}}}{2}\gamma_5\langle0|[\delta(\omega-\overline{\mathcal{P}})\chi_n(y_-)]|\Lambda X\rangle\langle \Lambda X|\bar{\chi}_n(0)|0\rangle\right]\,,\nn\\
&{S}^i_{\perp}\mathcal{G}^T_{\Lambda/q}(z,s)=\frac{1}{P^{\Lambda}_-}\int\frac{\md y_-}{4\pi} \, e^{ ik_+y_-/2} \sum_X \frac{1}{2N_c}\notag\\
&\hspace{6em} \times\mathrm{tr}\bigg[\frac{\slashed{\bar{n}}}{2} \gamma^i_{\perp}\gamma_5 \langle0|[\delta(\omega-\overline{\mathcal{P}})\chi_n(y_-)]|\Lambda X\rangle \langle \Lambda X|\bar{\chi}_n(0)|0\rangle \bigg]\,.
\end{align}
In the definitions above, $\overline{\mathcal{P}}$ denotes the momentum operator for the jet constituents, constraining the large light cone momentum component of the jet. The subsequent Fourier transform makes the FJF differential in the jet invariant mass, $ s=k_+ \omega$.

In the perturbative regime where the jet invariant mass is large ($s \gg \Lambda_{\text{QCD}}^2$), the FJF can be factorized further by matching it onto the non-perturbative FFs via the standard operator product expansion. In Laplace space, this matching takes the form as
\begin{align}\label{eq:conv}
{\mathcal{G}}^{\mathcal{P}}_{\Lambda/q}\left(z,\frac{u}{Q^2},\mu\right)=\sum_{j}\int_z^1\frac{\md x}{x} \, {\mathcal{J}}^{\mathcal{P}}_{jq}\left(\frac{z}{x},\frac{u}{Q^2},\mu\right) \mathcal{D}^{\mathcal{P}}_{\Lambda/j}(x, \mu)\,,
\end{align}
where the convolution involves the perturbatively calculable matching coefficients, $\mathcal{J}^{\mathcal{P}}_{jq}$, and the non-perturbative FFs, $\mathcal{D}^{\mathcal{P}}_{\Lambda/j}$, defined in \eqref{eq:FFs}. The sum over $j$ extends over all partons, including gluons and all light quark flavors.

The perturbative matching coefficients, ${\mathcal{J}}^{\mathcal{P}}_{jq}$, are expanded in powers of the strong coupling constant, and the NLO expressions are presented below. For brevity, we define $L_u=\ln(Q^2/\mu^2 u)$. The NLO bare matching coefficients read
\begin{align}\label{eq:matching_coefficients}
&{\mathcal{J}}^{U}_{q q}\left(x, \frac{u}{Q^2} \right) 
=\delta(1-x)+\frac{\alpha_s}{4\pi} \left[\delta(1-x)C_F\left(\frac{4}{\epsilon^2}+\frac{1}{\epsilon}\left(3-4 L_u\right)-3 L_u+2 L_u^2\right)\right. \nn \\ 
&\hspace{1cm}\left.+2(1-x) C_F+P_{q q}^{U(0)}(x) \ln x+2C_F \left(1+x^2\right)\left[\frac{\ln (1-x)}{1-x}\right]_{+}+\left(-\frac{1}{\epsilon}+L_u\right) P_{q q}^{U(0)}(x)\right]\,, \nn \\
&{\mathcal{J}}^{U}_{g q}\left(x, \frac{u}{Q^2} \right) 
=\frac{\alpha_s}{4\pi} \left[\left(-\frac{1}{\epsilon}+L_u\right) P_{g q}^{U(0)}(x)+ P_{g q}^{U(0)}(x) \ln (x(1-x))+2 x C_F \right]\,, \nn \\
&{\mathcal{J}}^{L}_{q q}\left(x, \frac{u}{Q^2} \right) 
={\mathcal{J}}^{U}_{q q}\left(x, \frac{u}{Q^2} \right)\,, \nn \\
&{\mathcal{J}}^{L}_{g q}\left(x, \frac{u}{Q^2} \right)  
=\frac{\alpha_s}{4\pi} \left[\left(-\frac{1}{\epsilon}+L_u\right)P_{g q}^{L(0)}(x)+ P_{g q}^{L(0)}(x) \ln (x(1-x))+(4 x-4)C_F\right]\,, \nn \\
&{\mathcal{J}}^{T}_{q q}\left(x, \frac{u}{Q^2} \right) 
=\delta(1-x)+\frac{\alpha_s}{4\pi} \left[\delta(1-x)C_F \left(\frac{4}{\epsilon^2}+\frac{1}{\epsilon}\left(3-4 L_u\right)-3 L_u+2 L_u^2\right) \right. \nn \\
&\hspace{1cm} \left. +P_{q q}^{T(0)}(x) \ln x+4x C_F \left[\frac{\ln (1-x)}{1-x}\right]_{+}+\left(-\frac{1}{\epsilon}+L_u\right) P_{q q}^{T(0)}(x)\right]\,.
\end{align}
Here $P_{jq}^{\mathcal{P}(0)}(z)$ denotes the LO Altarelli–Parisi splitting function with different polarization states, which are given by \cite{Kang:2023elg}
\begin{align}
P_{q q}^{U(0)}(z) & =2 C_F\left[\frac{1+z^2}{(1-z)_{+}}+\frac{3}{2} \delta(1-z)\right]\,, \quad\quad
P_{g q}^{U(0)}(z)  =2 C_F\left[\frac{z^2-2z+2}{z}\right]\,, \nn \\
P_{q q}^{L(0)}(z) & =2 C_F\left[\frac{1+z^2}{(1-z)_{+}}+\frac{3}{2} \delta(1-z)\right]\,, \quad\quad
P_{g q}^{L(0)}(z)  =2 C_F(2-z)\,, \nn \\
P_{q q}^{T(0)}(z) & =2 C_F\left[\frac{2 z}{(1-z)_{+}}+\frac{3}{2} \delta(1-z)\right]\,.
\end{align}
We note the absence of a $\mathcal{J}_{gq}^{T}$  term, which reflects the vanishing of the quark-to-gluon transversity splitting function at this order. After performing renormalization in the $\overline{\text{MS}}$ scheme, we obtain the final renormalized matching coefficients. The full details of this calculation are presented in appendix~\ref{app:FJFs}.

\subsection{Renormalization group evolution and resummation}\label{sec:Resummation}

In the dijet limit, the cross section contains large logarithms of the thrust variable $\tau$, which must be resummed to all orders in perturbation theory. This is achieved by solving the renormalization group equations (RGEs) for the hard, jet (FJF), and soft functions, and evolving them from their natural scales to a common factorization scale $\mu$. This procedure is most conveniently carried out in Laplace space.
Each component $F \in \{H, S, \mathcal{G}\}$ of the factorized cross section obeys a multiplicative RGE of the form
\begin{equation}
    \frac{\mathrm{d}}{\mathrm{d}\ln\mu} F(\mu) = \Gamma_F[\alpha_s(\mu)] F(\mu)\,,
\end{equation}
where $\Gamma_F$ is the corresponding anomalous dimension. The anomalous dimensions are composed of a universal cusp part, $\gamma^{\text{cusp}}$, and a non-cusp part, $\gamma^F$
\begin{align}
\Gamma_H &= 2C_F\,\gamma^{\text{cusp}}(\alpha_s)\ln\frac{Q^2}{\mu^2} + 2\gamma^H(\alpha_s)\,, \\
\Gamma_{\mathcal{G}} &=- 2C_F\,\gamma^{\text{cusp}}(\alpha_s)\ln\frac{Q^2}{u \mu^2}- 2\gamma^\mathcal{G}(\alpha_s)\,, \\
\Gamma_S &= 2C_F\,\gamma^{\text{cusp}}(\alpha_s)\ln\frac{Q^2}{u^2\mu^2} - 2\gamma^S(\alpha_s)\,.
\end{align}
Here the RGE invariance requires that $\Gamma_H+2\,\Gamma_{\mathcal{G}}+\Gamma_S=0$. Solving these equations from a characteristic scale to the final scale $\mu$ introduces evolution factors. For example, the resummed hard function is given by
\begin{align}
H(Q^2,\mu)=H(Q^2,\mu_h)\exp\left[4C_FS(\mu_h,\mu)-2A_H(\mu_h,\mu)\right]\left(\frac{Q^2}{\mu_h^2}\right)^{-2C_F A_{\text{cusp}}(\mu_h,\mu)}\,.
\end{align}
The evolution kernels $S$ and $A_i$ are defined as integrals of the anomalous dimensions over the running coupling $\alpha_s(\mu)$
\begin{align}
    S(\nu,\mu) &= -\int_{\alpha_s(\nu)}^{\alpha_s(\mu)}\mathrm{d}\alpha\frac{\gamma^{\text{cusp}}(\alpha)}{\beta(\alpha)}\int_{\alpha_s(\nu)}^{\alpha}\frac{\mathrm{d}\alpha^{\prime}}{\beta(\alpha^{\prime})}\,,\\
    A_i(\nu,\mu) &= -\int_{\alpha_s(\nu)}^{\alpha_s(\mu)}\mathrm{d}\alpha\frac{\gamma^i(\alpha)}{\beta(\alpha)}\,,\qquad i \in \{\text{cusp}, H, \mathcal{G}, S\}\,.
\end{align}
The expressions for the anomalous dimensions and evolution kernels required for NNLL resummation are collected in appendix~\ref{app:func}.

Combining the evolution of all components from their natural scales ($\mu_h \sim Q$, $\mu_J \sim \sqrt{\tau} Q$, $\mu_s \sim \tau Q$) and performing the inverse Laplace transform yields the resummed cross section integrated up to a value $\tau_{\text{cut}}$
\begin{align}\label{eq:resum}
        \frac{\md \sigma^{\mathcal{P}}\left(\tau_{\text{cut}}\right)}{ \md z_1 \, \md z_2\, \md \Omega}  =&\int_0^{\tau_{\text{cut}}}\md\tau\,\frac{\md \sigma^{\mathcal{P}}}{\md \tau \,  \md z_1 \, \md z_2\, \md \Omega}  \\
        =&
        \frac{\md \sigma^{\mathcal{P}}_0}{\md\Omega}\exp\left[4C_FS(\mu_h,\mu_J)+4C_FS(\mu_s,\mu_J)-2A_H(\mu_h,\mu_s)+4A_J(\mu_J,\mu_s)\right]\,\notag\\
        & \times \left(\frac{Q^2}{\mu_h^2}\right)^{-2C_F A_{\text{cusp}}(\mu_h,\mu_J)}H(Q^2,\mu_h)\,\widetilde{S}_{T}\left(\partial_\eta,\mu_s\right)\,\notag\\
        & \times \sum_q {e_q^2}\,\widetilde{\mathcal{G}}^{\mathcal{P}}_{{\Lambda}/{q} }\left(z_1,\ln\frac{\mu_sQ}{\mu_J^2}+\partial_\eta,\mu_J\right)\widetilde{\mathcal{G}}^{\mathcal{P}}_{{\bar{\Lambda}}/{\bar{q}} }\left(z_2,\ln\frac{\mu_sQ}{\mu_J^2}+\partial_\eta,\mu_J\right) 
        \,\notag\\
        & \times \left(\frac{\tau_{\text{cut}} Q}{\mu_s}\right)^\eta\frac{e^{-\gamma_E\eta}}{\Gamma(1+\eta)}\Bigg|_{\eta=4 C_F A_{\text{cusp}}(\mu_J,\,\mu_s)}\,. \notag
\end{align}
Here, the derivative notation, e.g., $\widetilde{S}_T(\partial_\eta, \mu_s)$, provides a compact analytical representation of the inverse Laplace transform, allowing for a direct calculation in momentum space~\cite{Becher:2014oda, Becher:2008cf, Becher:2006mr}. An alternative approach is to perform the resummation in Laplace (Mellin) space and then numerically invert the result using a complex contour integration \cite{Catani:1992ua, Sterman:1986aj}. A detailed comparison of these numerical approaches for the thrust resummation can be found in Ref.~\cite{Monni:2011gb}. While both methods are formally equivalent to the logarithmic accuracy considered in this work, they can differ by power-suppressed terms arising from the treatment of the Landau pole in the QCD coupling constant. In our numerical implementation, the hard, jet, and soft scales remain in the perturbative domain, and we therefore do not introduce any non-perturbative prescription for the Landau pole.

With the resummed polarized cross section at hand, we can directly relate it to the density matrix formalism introduced in Sec.~\ref{subsec:FDM}. The $\Lambda\bar{\Lambda}$ production spin density matrix retains the same structure as in Eq.~\eqref{eq:density_matrix}, allowing for the calculation of spin correlations from the resummed polarized cross sections using the same projection methods as given in the previous section.

\section{Bell nonlocality and decoherence}\label{sec:decoherence}

Having established the spin correlations within both the parton model and the resummation framework, we now adopt a quantum information perspective on these high-energy processes. In recent years, the language of quantum information theory has proven invaluable for reinterpreting fundamental aspects of QCD \cite{Neill:2018uqw, Datta:2024hpn, vonKuk:2025kbv, Qi:2025onf}. A central concept in this context is decoherence, which describes the loss of quantum correlations due to interactions with an environment. For QCD, fragmentation and hadronization provide such an environment, potentially reducing the entanglement generated during the initial hard partonic scattering stage \cite{Gong:2021bcp, Datta:2024hpn}. For example, as shown in \cite{Aoude:2025ovu}, the emission of unresolved radiation can reduce the spin entanglement of fermion pairs, acting analogously to an environment that traces out unmeasured degrees of freedom. These considerations motivate viewing QCD processes as open quantum systems: while the hard interaction may produce an entangled quark–antiquark pair, the subsequent parton shower and hadronization stages interact with this system and induce decoherence.

Motivated by these insights, we investigate how QCD-induced decoherence affects the spin entanglement of $\Lambda\bar{\Lambda}$ pairs produced in $e^+e^-$ annihilation. In particular, if the initial hard process generates a maximally entangled quark and anti-quark pair, it is natural to ask whether—and to what extent—this entanglement survives the subsequent fragmentation and hadronization stages. To explore this question quantitatively, we model the spin state of the final-state $\Lambda\bar{\Lambda}$ pair as a bipartite qubit system, described by a $4\times4$ spin density matrix $\hat{\rho}_{\Lambda\bar{\Lambda}}$.\footnote{Modeling the $\Lambda\bar{\Lambda}$ pair as a bipartite qubit system is a well-motivated choice for spin-1/2 particles within the collinear factorization framework. Our treatment neglects potential spin-momentum correlations described by the standard transverse momentum dependent factorization framework \cite{Boussarie:2023izj}, the study of which is an interesting avenue for future work.} A general two-qubit density matrix for particles $A$ and $B$ can be decomposed onto a basis of tensor products of Pauli matrices
\begin{align}\label{eq:bipartite density matrix}
\hat{\rho}_{AB} = \frac{1}{4} \left( \hat{I}^A \otimes \hat{I}^B + \sum_{i} P_i^A\, \hat{\sigma}^A_i \otimes \hat{I}^B + \sum_{j} P_j^B\, \hat{I}^A \otimes \hat{\sigma}^B_j + \sum_{i,j} C_{ij}\, \hat{\sigma}^A_i \otimes \hat{\sigma}^B_j \right)\,,
\end{align}
where $P_i^A$ and $P_j^B$ are the components of the polarization (or Bloch) vectors of particles $A$ and $B$, respectively. These vectors encode the individual spin polarizations along the $i$-th and $j$-th axes in each particle’s helicity frame. The tensor $C_{ij} = \text{Tr}\left( \hat{\rho}\, \hat{\sigma}^A_i \otimes \hat{\sigma}^B_j \right)$ captures the spin correlations between the two subsystems along different directions \cite{Baumgart:2012ay, CMS:2019nrx}. As in Eq.~\eqref{eq:density_matrix_partonic}, the Pauli matrices $\hat{\sigma}^A_i$ and $\hat{\sigma}^B_j$ represent the spin operators of particles $A$ and $B$ projected along the $i$-th and $j$-th coordinate axes in their respective helicity frames.

To illustrate this decomposition, we first consider the process at the parton level. The production density matrix for the $q\bar{q}$ pair, derived from Eq.~\eqref{eq:density_matrix_partonic}, is normalized to have unit trace and can be written as
\begin{align}\label{eq:density matrix qqbar}
\hat{\rho}_{q\bar{q}} = \frac{1}{4} \left( \hat{I}^{q} \otimes \hat{I}^{\bar{q}}
+ \frac{\sin^2\theta}{1+\cos^2\theta}\, \hat{\sigma}^{q}_{x} \otimes \hat{\sigma}^{\bar{q}}_{\bar{x}}
+ \frac{\sin^2\theta}{1+\cos^2\theta}\, \hat{\sigma}^{q}_{y} \otimes \hat{\sigma}^{\bar{q}}_{\bar{y}}
- \hat{\sigma}^{q}_{z} \otimes \hat{\sigma}^{\bar{q}}_{\bar{z}} \right)\,.
\end{align}
By comparing this expression with the general form of the bipartite density matrix in Eq.~\eqref{eq:bipartite density matrix}, we can directly identify the partonic spin correlation matrix
\begin{align}
C^{q\bar{q}}_{ij} = \text{diag} \left( \frac{\sin^2\theta}{1+\cos^2\theta}, \frac{\sin^2\theta}{1+\cos^2\theta}, -1 \right)\,.
\end{align}
We also find that the single-parton polarization vectors are zero, $P_i^A = P_j^B = 0$, as expected from parity conservation.

We now turn to the physical $\Lambda\bar{\Lambda}$ system. Its production density matrix, which includes hadronization effects, is given by Eq.~\eqref{eq:density_matrix}. Upon normalization, it can be written as
\begin{align}\label{eq:density matrix LLbar}
\hat{\rho}_{\Lambda\bar{\Lambda}} = \frac{1}{4} \left( \hat{I}^{\Lambda} \otimes \hat{I}^{\bar{\Lambda}}
+\frac{\md \sigma^T}{\md \sigma^U} \, \hat{\sigma}^{\Lambda}_{x} \otimes \hat{\sigma}^{\bar{\Lambda}}_{\bar{x}}
+ \frac{\md \sigma^T}{\md \sigma^U}\, \hat{\sigma}^{\Lambda}_{y} \otimes \hat{\sigma}^{\bar{\Lambda}}_{\bar{y}}
-\frac{\md \sigma^L}{\md \sigma^U} \hat{\sigma}^{\Lambda}_{z} \otimes \hat{\sigma}^{\bar{\Lambda}}_{\bar{z}} \right)\,.
\end{align}
By comparing this with the general form in Eq.~\eqref{eq:bipartite density matrix}, we can directly identify the correlation matrix for the hyperon pair
\begin{align}
C^{\Lambda\bar{\Lambda}}_{ij} = \text{diag} \left( \frac{\md \sigma^T}{\md \sigma^U}, \frac{\md \sigma^T}{\md \sigma^U}, -\frac{\md \sigma^L}{\md \sigma^U} \right)\,.
\end{align}
To probe the quantum nature of these correlations, we employ the Clauser-Horne-Shimony-Holt (CHSH) inequality, the standard Bell test for two-qubit systems. This inequality establishes a universal bound on the strength of correlations permissible within any local hidden-variable theory~\cite{Clauser:1969ny}:
\begin{align}
\left| \bm{a}_1^\mathrm{T} \cdot \bm{C} \cdot \left( \bm{b}_1+ \bm{b}_2 \right) + \bm{a}_2^\mathrm{T}  \cdot \bm{C} \cdot \left( \bm{b}_1 -\bm{b}_2 \right) \right| \leq 2\,,
\end{align}
where $\bm{a}_i$ and $\bm{b}_i$ are unit vectors defining the spin measurement axes, and $\bm{C}$ is the correlation matrix from Eq.~\eqref{eq:bipartite density matrix}.  A violation of this bound signals an incompatibility with classical realism and occurs if and only if the sum of the two largest eigenvalues of $\bm{C}^\mathrm{T}\bm{C}$ exceeds one~\cite{Horodecki:1995nsk}. For practical analyses, specific choices for the measurement axes lead to a simpler, sufficient criterion for Bell violation, expressed in terms of two Bell variables~\cite{Severi:2021cnj}
\begin{align}
\mathcal{B}_\pm = C_{xx} \pm C_{yy}\,,
\end{align}
with a violation of the CHSH-Bell inequality implied if
\begin{align}
\max \left( |\mathcal{B}_+|,\, |\mathcal{B}_-| \right) > \sqrt{2}\,.
\end{align}

To establish a baseline, we first consider the ideal partonic system. For the $q\bar{q}$ final state, the Bell variables are
\begin{align}\label{eq:bell_parton}
\mathcal{B}_+^{q\bar{q}} = \frac{2\,\md\sigma_0^T}{\md\sigma_0^U}= \frac{2\sin^2\theta}{1+\cos^2\theta}\,, \qquad \mathcal{B}_-^{q\bar{q}} = 0\,.
\end{align}
In the central scattering region ($\theta=\pi/2$), $\mathcal{B}_+^{q\bar{q}}$ reaches its algebraic maximum of $2$. This corresponds to the correlation matrix $C_{ij}^{q\bar q} = \text{diag}(1,1,-1)$ of the maximally entangled Bell state.

For the hadronic $\Lambda\bar{\Lambda}$ final state, where spin correlations are modified by fragmentation and hadronization, the Bell variables are determined as
\begin{align}\label{eq:bell_hadron}
\mathcal{B}_+^{\Lambda\bar{\Lambda}} = \frac{2\,\md\sigma^T}{\md\sigma^U}\,, \qquad \mathcal{B}_-^{\Lambda\bar{\Lambda}} = 0\,.
\end{align}
This result establishes transverse spin correlation as a direct, quantitative probe of entanglement survival. A suppression of $\mathcal{B}_+$ relative to its partonic value would signal decoherence effects induced by the fragmentation process. The phenomenological implications of this analysis are discussed in Sec.~\ref{sec:Pheno}, where we evaluate $\mathcal{B}_+^{\Lambda\bar{\Lambda}}$ using different hadronization models.

\section{Numerical results}\label{sec:Pheno}

In this section, we present our numerical predictions in $e^+e^- \to \Lambda\bar{\Lambda}X$. The results are evaluated as a function of the momentum fractions $z_1$ and $z_2$ carried by the $\Lambda$ and $\bar{\Lambda}$, respectively. To select the dijet topology relevant for experiments such as Belle II, we impose a thrust cut of $\tau < \tau_{\text{cut}} = 0.2$ \cite{Belle:2008fdv, Belle:2011cur, Belle:2018ttu}.

\subsection{Resummed cross section}\label{sec:xsec}

We now present our numerical results based on the resummation formalism derived in Sec.~\ref{sec:Resummation}. The cross sections are evaluated at two distinct center-of-mass energies: $Q=10.58$~\text{GeV} and $Q=100$~\text{GeV}, corresponding to the Belle II and LEP, respectively. For our central predictions, we choose the following canonical scales for the hard, jet, and soft functions:
\begin{align}
    \mu_h = Q, \quad \mu_J= Q\sqrt{\tau_{\text{cut}}}, \quad \mu_s = Q\tau_{\text{cut}}.
\end{align}

\begin{figure}[t!]
    \centering
    \begin{minipage}[t]{0.48\textwidth}
        \centering
        \subfloat[Unpolarized]{\includegraphics[width=0.9\linewidth]{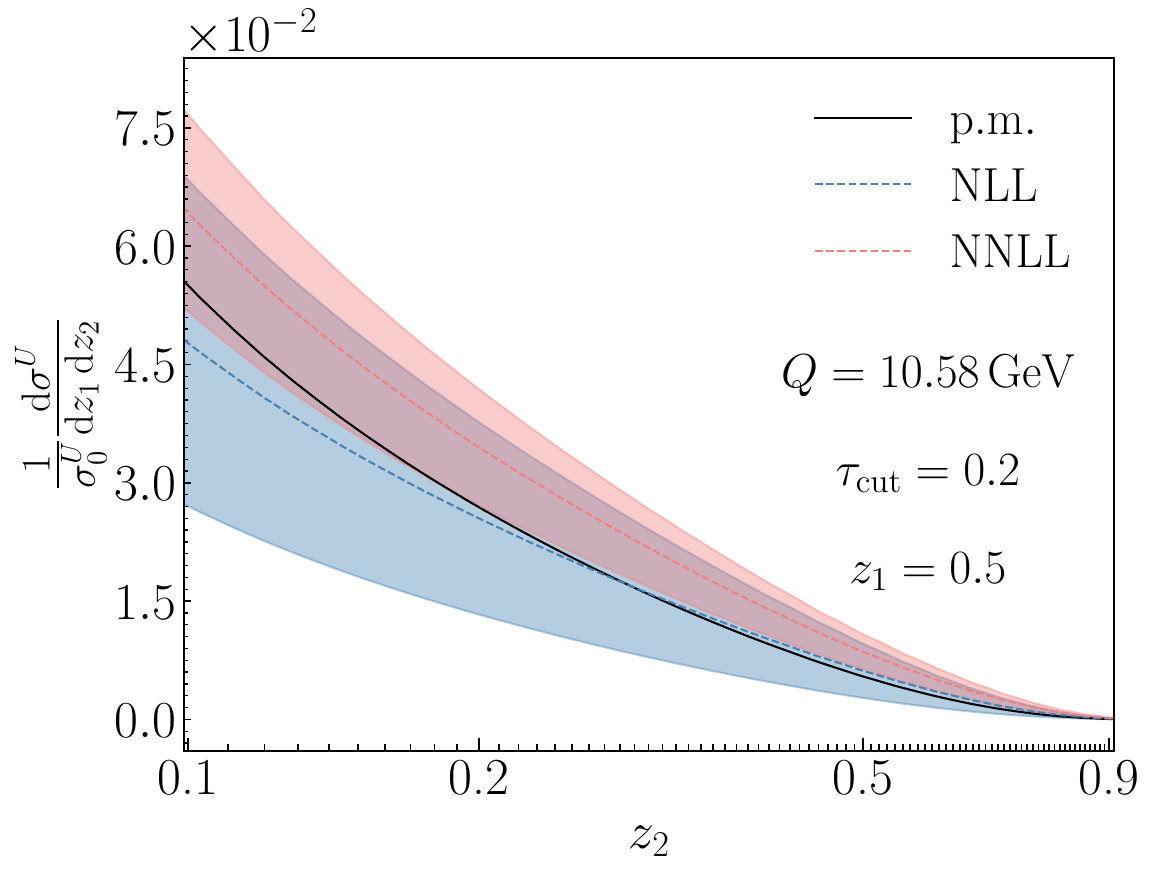}\label{subfig:unpol_10gev}}\\[1ex]
        \subfloat[Longitudinally polarized]{\includegraphics[width=0.9\linewidth]{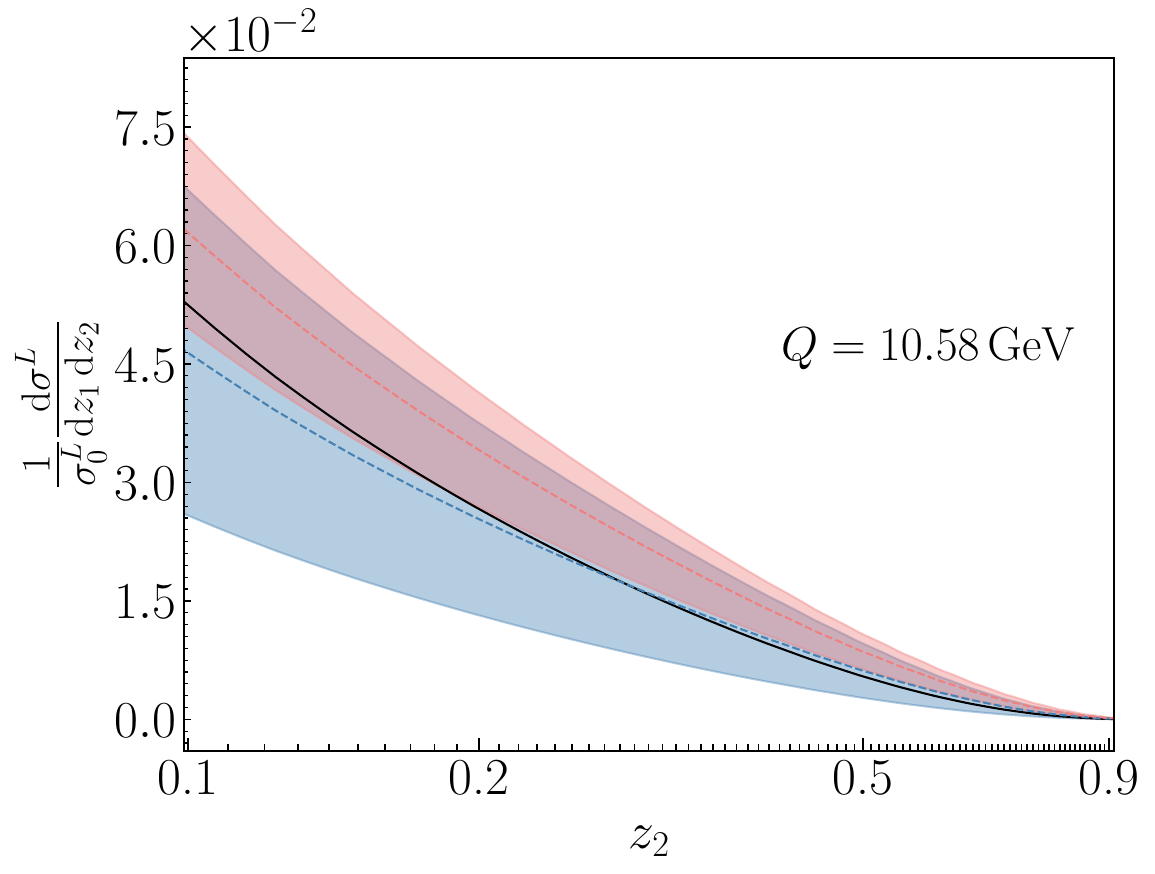}\label{subfig:lpol_10gev}}\\[1ex]
        \subfloat[Transversely polarized]{\includegraphics[width=0.9\linewidth]{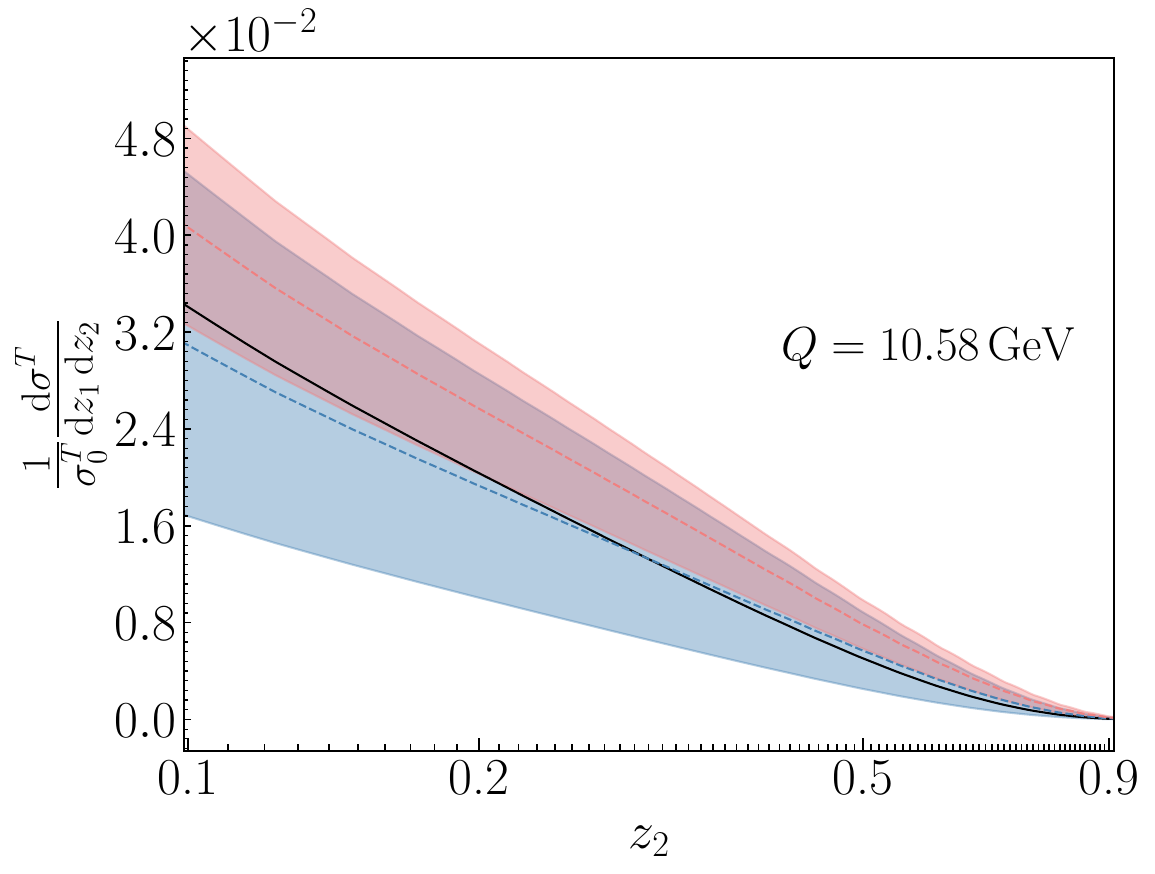}\label{subfig:tpol_10gev_v}}\\[1ex]
    \end{minipage}%
    \hfill
    \begin{minipage}[t]{0.48\textwidth}
        \centering
        \subfloat[Unpolarized]{\includegraphics[width=0.9\linewidth]{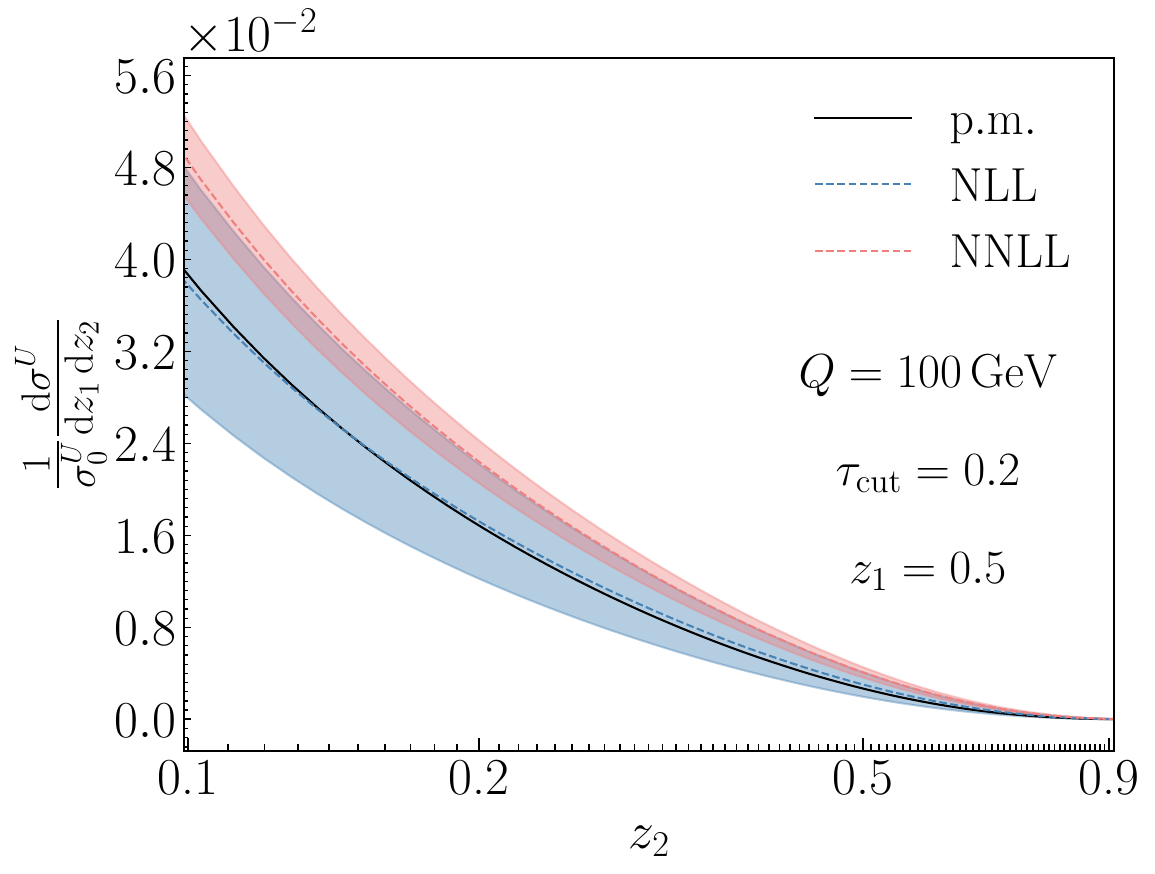}\label{subfig:unpol_100gev}}\\[1ex]
        \subfloat[Longitudinally polarized]{\includegraphics[width=0.9\linewidth]{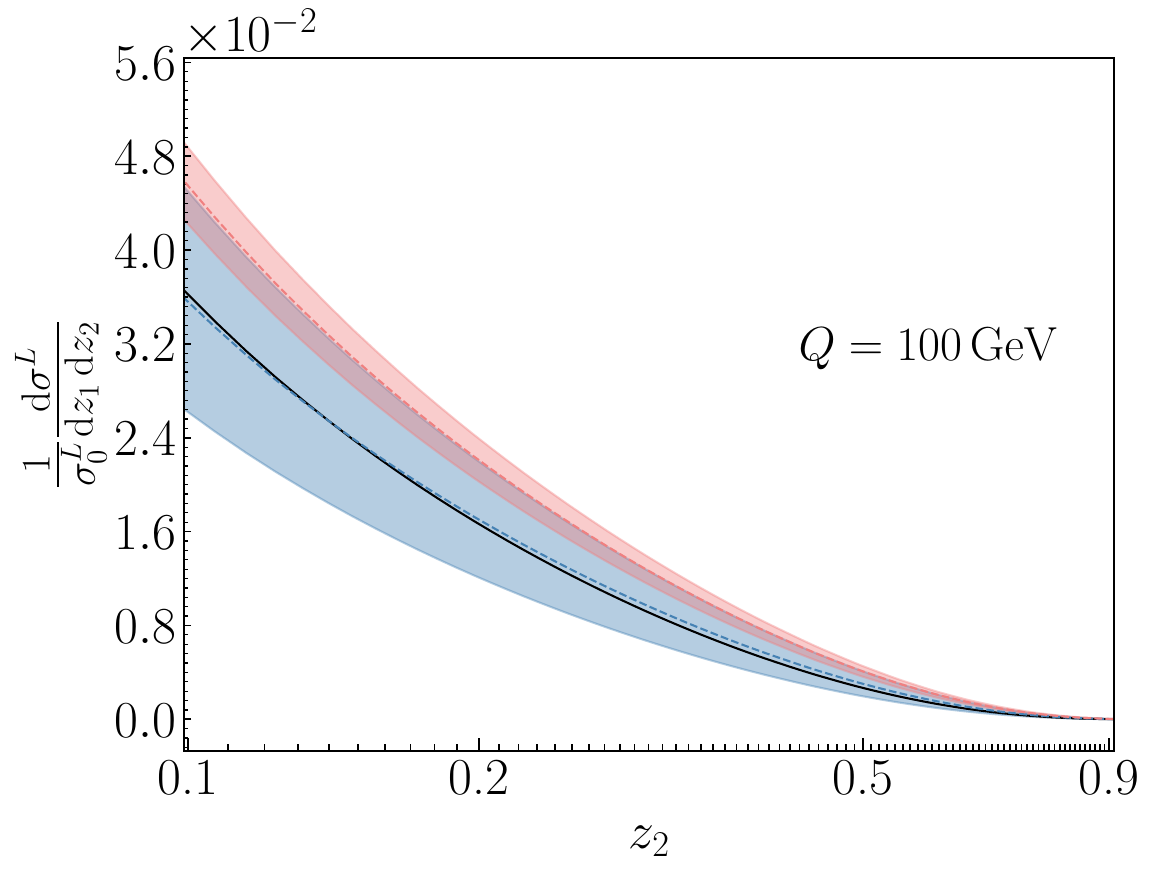}\label{subfig:lpol_100gev}}\\[1ex]
        \subfloat[Transversely polarized]{\includegraphics[width=0.9\linewidth]{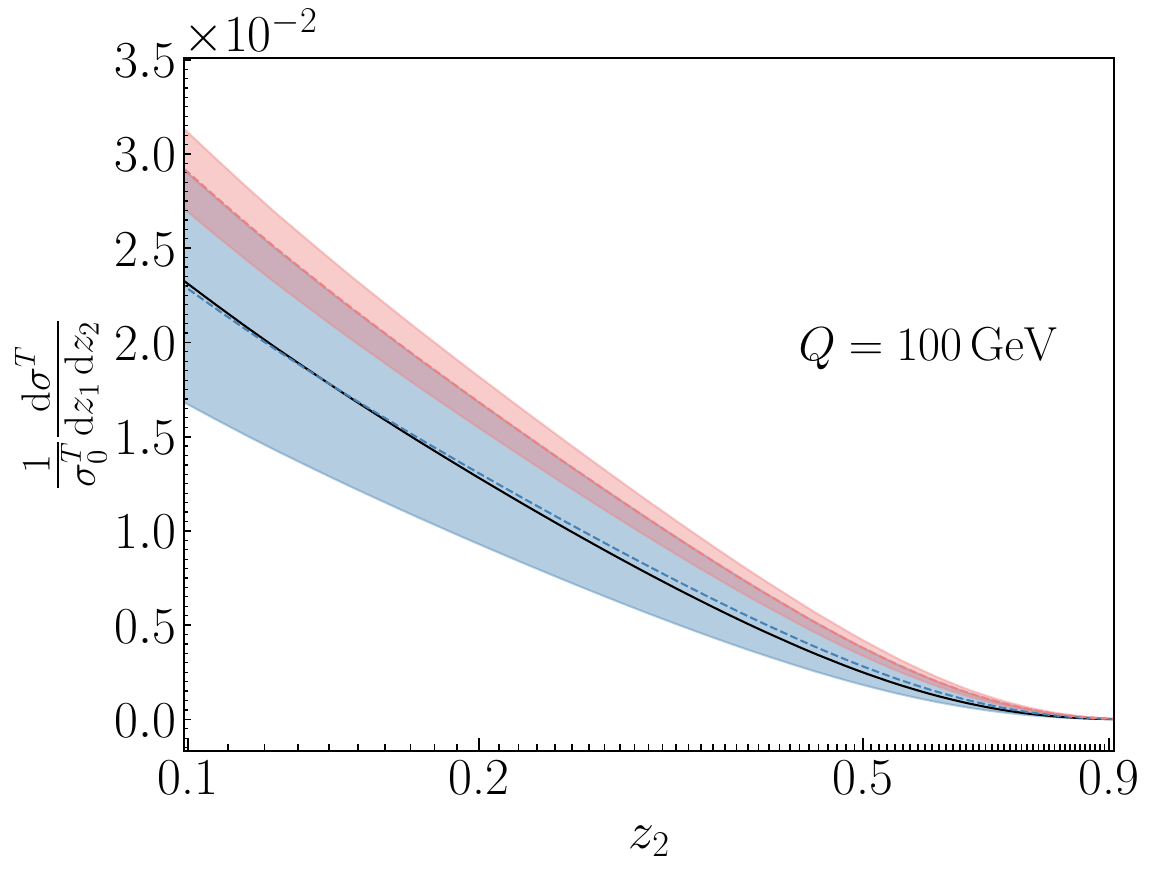}\label{subfig:lpol_100gev_v}}\\[1ex]
    \end{minipage}
    \caption{Resummed cross sections for $\Lambda\bar{\Lambda}$ production integrated up to $\tau_{\text{cut}}=0.2$, showing the unpolarized (a, d), longitudinally polarized (b, e), and transversely polarized (c, f) components. The left panels show results for $Q=10.58$~GeV, while the right panels are for $Q=100$~GeV. The wider, blue bands correspond to the NLL prediction, and the narrower, red bands represent the NNLL result. The black line represents the parton model (p.m.) result. The bands indicate the theoretical uncertainty from scale variations.}
    \label{fig:resum_xsec}
\end{figure}

For the non-perturbative inputs to our calculation, we employ the de Florian-Stratmann-Vogelsang (DSV) parameterization for the unpolarized $\Lambda$ FF, $\mathcal{D}^U(z,\mu)$~\cite{deFlorian:1997zj}. As the corresponding polarized FFs are not well-constrained by experimental data, we utilize theoretical positivity bounds to define their maximal contribution~\cite{Soffer:1994ww, Vogelsang:1997ak}. These constraints, evaluated at an initial scale of $\mu_0^2=2.5$~GeV$^2$, are given by
\begin{align}\label{eq:positivity}
|\mathcal{D}^L(z,\mu_0)| \leq \mathcal{D}^U(z,\mu_0)\,, \qquad |\mathcal{D}^T(z,\mu_0)| \leq \frac{1}{2}\left[\mathcal{D}^U(z,\mu_0) + \mathcal{D}^L(z,\mu_0)\right]\,.
\end{align}
The first inequality ensures that the helicity-dependent FF does not exceed the unpolarized one, while the second is a consequence of the Soffer bound, which guarantees the positivity of the transversity distribution. While these bounds are laid out at the initial scale, the DGLAP evolution does not induce any violations \cite{Vogelsang:1997ak}.

Crucially, by saturating these bounds, we can establish a theoretical upper limit on the spin correlations. This allows us to determine the maximum possible violation of the Bell inequality permitted by the fundamental structure of QCD, providing a valuable benchmark for interpreting experimental data and quantifying the effects of decoherence.

Figure~\ref{fig:resum_xsec} displays the resummed cross sections for unpolarized, longitudinally polarized, and transversely polarized $\Lambda\bar\Lambda$ production. The left column shows the results at $Q=10.58$~GeV, while the right column corresponds to $Q=100$~GeV. In each panel, the solid black line shows the parton model prediction including the DGLAP evolution, evaluated at a scale $\mu=Q$. The blue bands represent the next-to-leading logarithmic (NLL) predictions, and the red bands show the NNLL results. The uncertainty bands are derived from varying the hard, jet, and soft scales by a factor of two around their central values. Two key features are apparent from the comparison. First, the significant reduction in the uncertainty from NLL to NNLL, combined with their substantial overlap, demonstrates the robust perturbative convergence of the resummation calculation. Second, the parton-model result is broadly consistent with the central value of the NLL prediction, especially at $Q=100$~GeV. This is expected, as the NLL resummation primarily modifies the evolution kernel while keeping the hard, soft, and jet functions at LO. In contrast, the NNLL calculation incorporates one-loop matching coefficients, two-loop single logarithmic and three-loop double logarithmic anomalous dimensions, which systematically shift the central value away from the parton-model result. Therefore, the improved convergence and reduced uncertainty demonstrate the NNLL prediction as the most reliable theoretical result.

\subsection{Spin correlations}\label{sec:correlations}

\begin{figure}[t]
    \centering
    \begin{minipage}[t]{0.48\textwidth}
        \centering
        \subfloat[Scenario 1]{\includegraphics[width=0.9\linewidth]{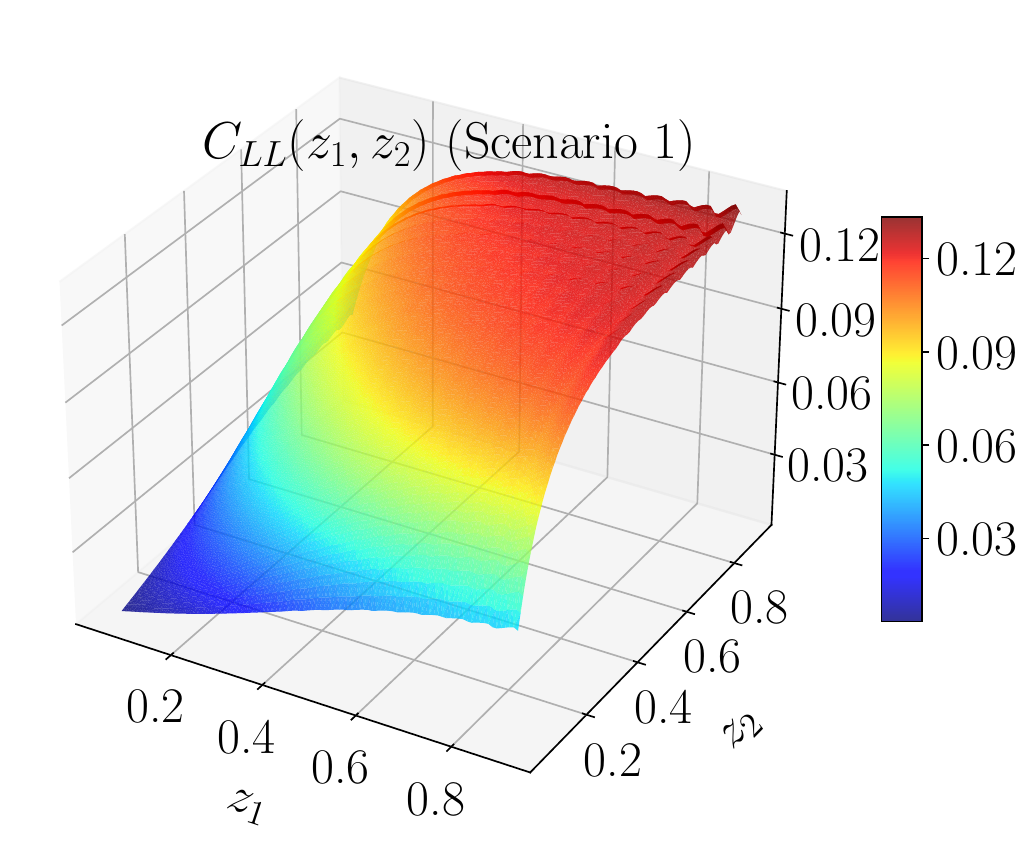}\label{fig:lpol3d_scenario1}}\\[1ex]
        \subfloat[Scenario 2]{\includegraphics[width=0.9\linewidth]{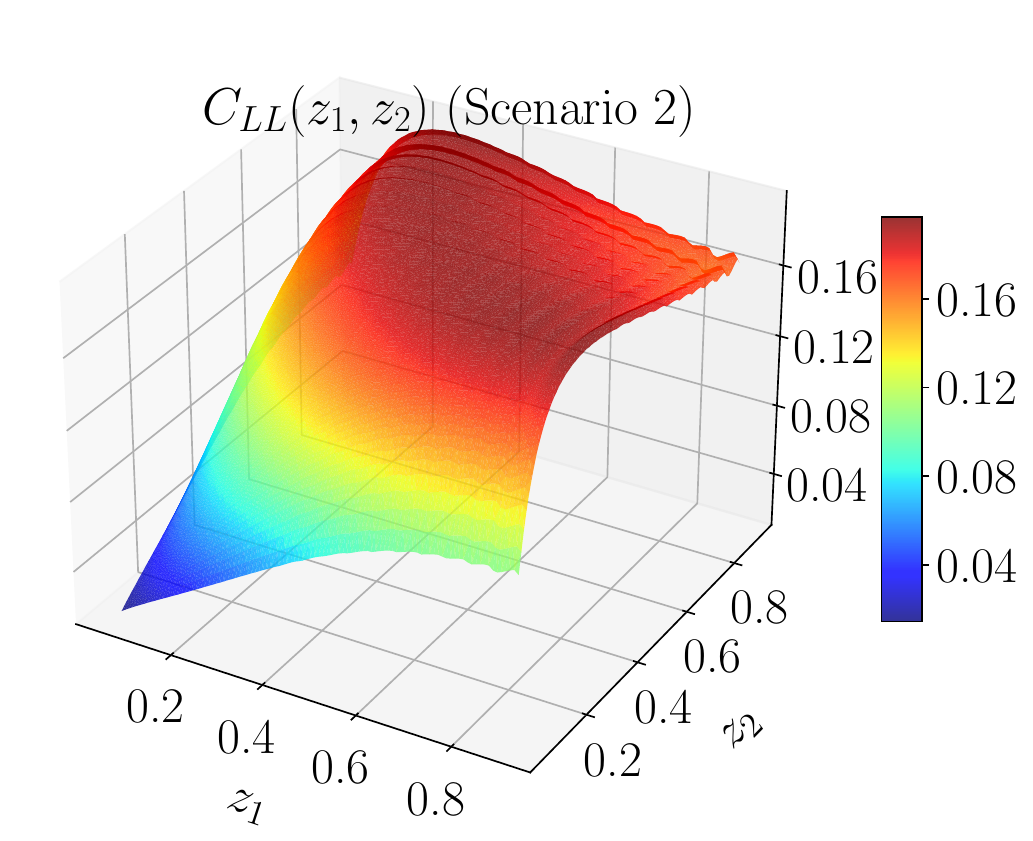}\label{fig:lpol3d_scenario2}}\\[1ex]
        \subfloat[Scenario 3]{\includegraphics[width=0.9\linewidth]{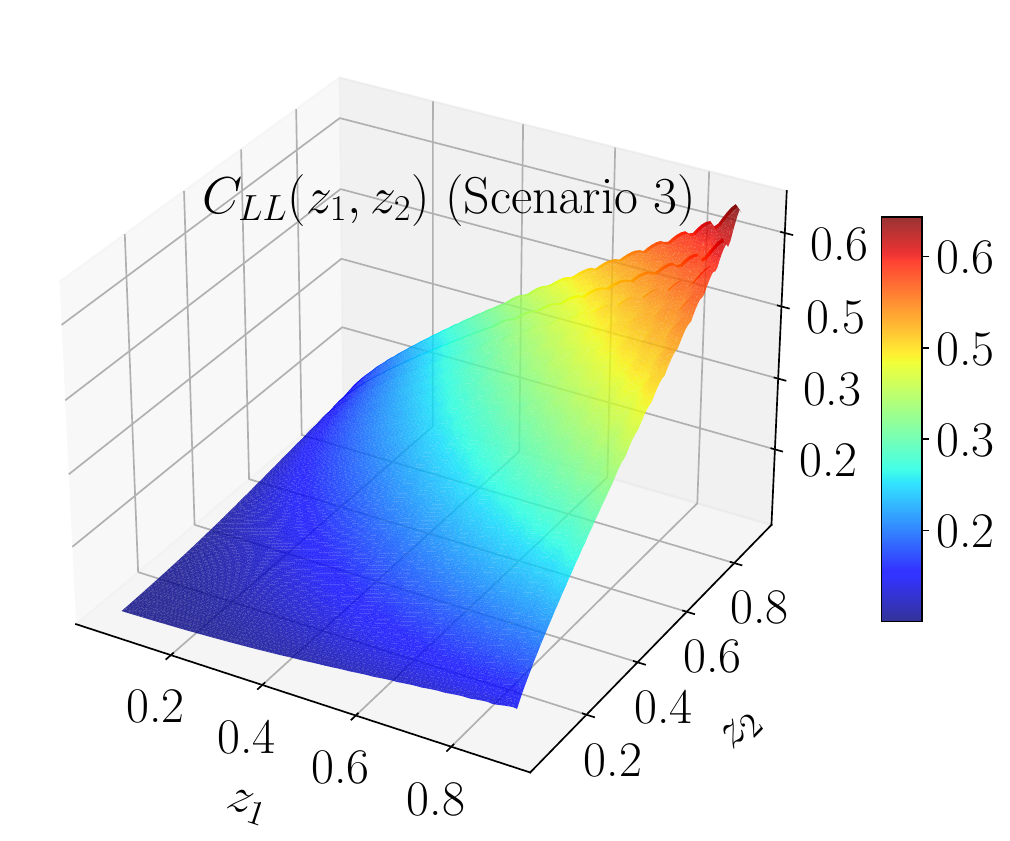}\label{fig:lpol3d_scenario3}}
        \caption*{Longitudinal spin correlation $C_{LL}$}
    \end{minipage}
    \hfill
    \begin{minipage}[t]{0.48\textwidth}
        \centering
        \subfloat[Scenario 1]{\includegraphics[width=0.9\linewidth]{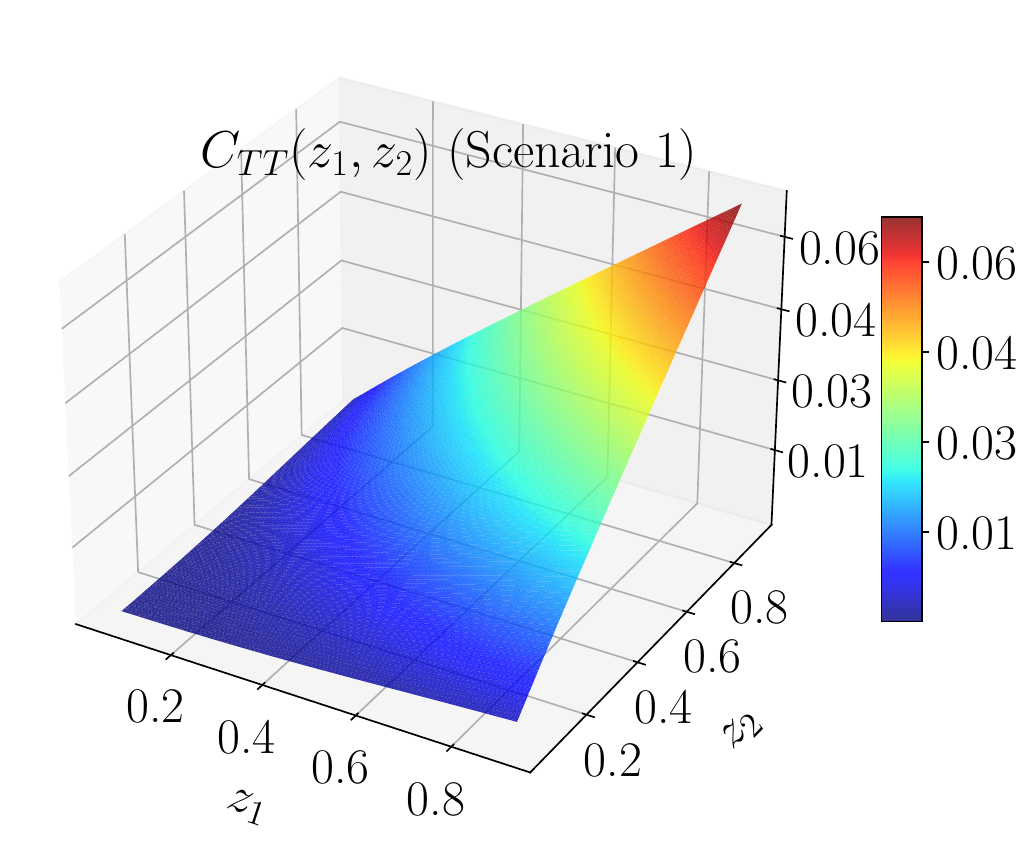}\label{fig:trans3d_scenario1}}\\[1ex]
        \subfloat[Scenario 2]{\includegraphics[width=0.9\linewidth]{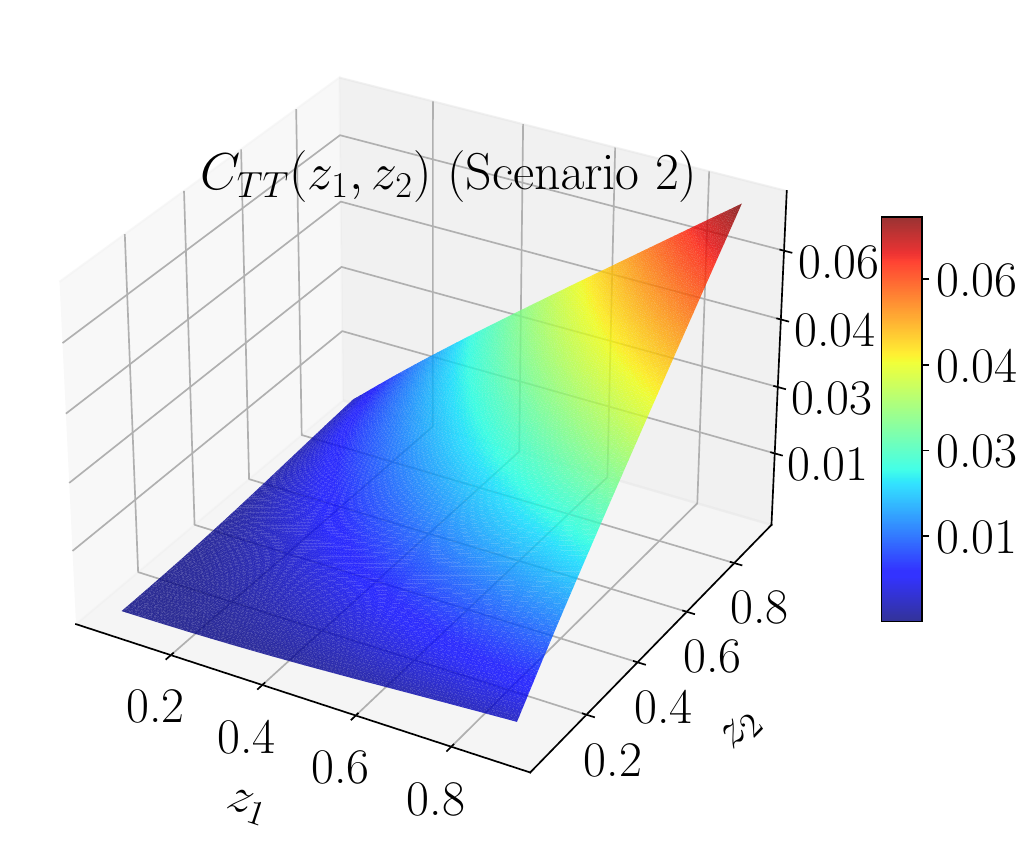}\label{fig:trans3d_scenario2}}\\[1ex]
        \subfloat[Scenario 3]{\includegraphics[width=0.9\linewidth]{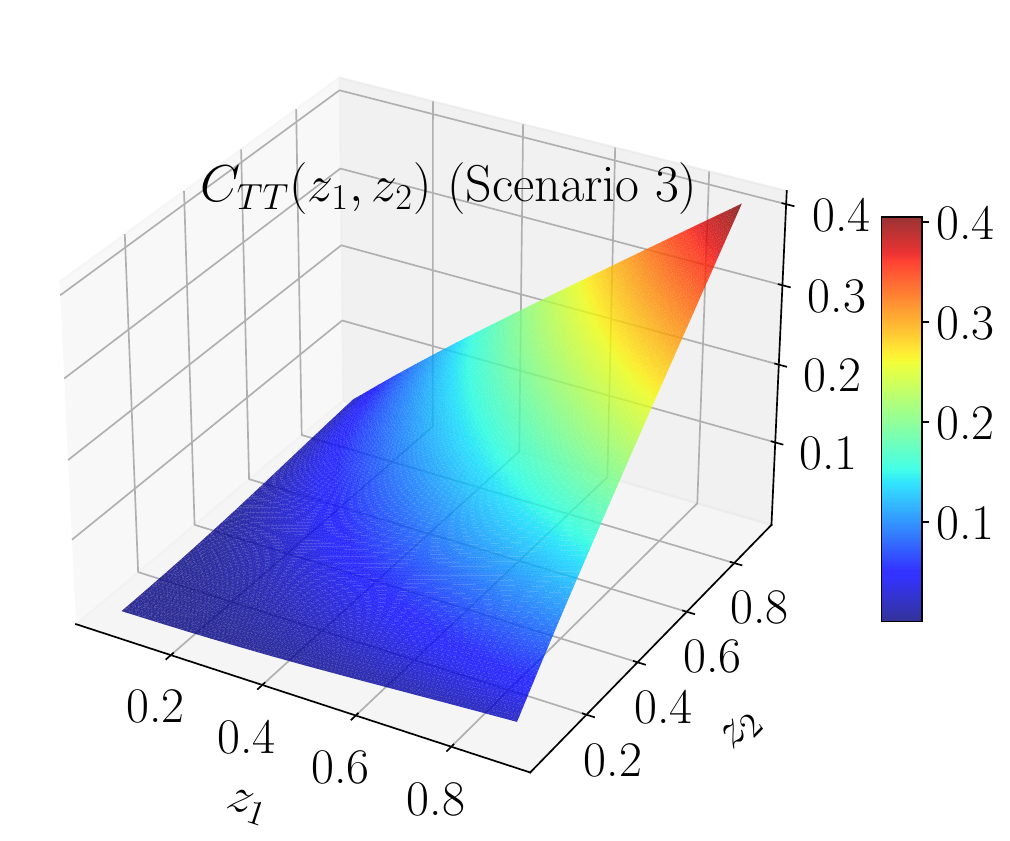}\label{fig:trans3d_scenario3}}
        \caption*{Transverse spin correlation $C_{TT}$}
    \end{minipage}%
        \caption{Contour plots of the spin correlations $C_{LL}$ (left column) and $C_{TT}$ (right column) as a function of the hyperon momentum fractions $z_1$ and $z_2$. The rows correspond to the three different models for the polarized FFs described in the text. All predictions are for $Q=10.58$~GeV with a thrust cut of $\tau < 0.2$. 
        }
    \label{fig:3d}
\end{figure}

In this section, we evaluate the spin correlations, $C_{LL}$ and $C_{TT}$, to explore their sensitivity to the flavor structure of the polarized FFs. Although several theoretical studies \cite{Burkardt:1993zh,deFlorian:1997zj, Boros:1998kc, Ma:1998pd, Chen:2016iey}  exist for longitudinally polarized FFs, the transverse sector remains largely uncharted. So far, only a handful of experimental data can be used to investigate polarized FFs. Therefore, theoretical studies rely heavily on physically motivated models. For the longitudinally polarized FFs, we adopt the DSV parameterization, which provides three scenarios. Following a similar strategy, we construct three corresponding models for the transversely polarized FFs:
\begin{itemize}
\item \textbf{Scenario 1 (Static quark model scenario)}: In the static quark model, the spin of the $\Lambda$ hyperon is carried by the $s$ quark, while $u$ and $d$ quarks completely do not contribute. We assume ${\cal D}^{T}_{s} (z, \mu_0) = z \, {\cal D}^{U}_s (z,\mu_0)$ and ${\cal D}^{T}_{u} (z, \mu_0) = {\cal D}^{T}_{d} (z, \mu_0) =0$ at the initial scale $\mu_0$.
\item \textbf{Scenario 2 (Burkardt-Jaffe scenario)}: This scenario assumes that $u$ and $d$ quarks contribute a small but negative component to the $\Lambda$ spin. At the initial scale, we assume ${\cal D}^{T}_{u/d} (z,\mu_0) = - 0.1 \, z \, {\cal D}^{U}_{u/d} (z,\mu_0)$ and ${\cal D}^{T}_{s} (z, \mu_0) = z \, {\cal D}^{U}_s (z,\mu_0)$.
\item \textbf{Scenario 3 (SU(3) symmetric scenario)}: The $u$, $d$, and $s$ quarks are assumed to contribute equally with positive polarized FFs, reflecting an approximate SU(3) flavor symmetry. We thus find ${\cal D}^{T}_{u/d/s} (z,\mu_0) = z \, {\cal D}^{U}_{u/d/s} (z,\mu_0)$.
\end{itemize}

Figure~\ref{fig:3d} displays the spin correlations for each of the three FF scenarios. The results reveal that the longitudinal correlation, $C_{LL}$, exhibits significant variation across all three models. In contrast, the transverse correlation, $C_{TT}$, shows similar results for the static quark model (Scenario 1) and Burkardt-Jaffe (Scenario 2) models, but differs by nearly an order of magnitude in the SU(3) symmetric scenario (Scenario 3). This behavior can be traced to the underlying flavor structure: the first two scenarios are both dominated by the strange quark's transversity, whereas the SU(3) symmetric scenario introduces large light-quark contributions. The power to distinguish between the first two scenarios is therefore provided by the longitudinal correlation, $C_{LL}$. Thus, the two observables offer complementary information, and their precise measurement at facilities like Belle II and the future electron-ion colliders \cite{Accardi:2012qut, AbdulKhalek:2021gbh, Anderle:2021wcy} will provide powerful new constraints on the spin and flavor dynamics of hadronization.

\subsection{Bell nonlocality}\label{sec:bellvar}

\begin{figure}[t]
    \centering
    \begin{minipage}[t]{0.48\textwidth}
        \centering
        \subfloat[positivity bound, $Q=10.58$~GeV, $\theta=\pi/2$]{\includegraphics[width=0.9\linewidth]{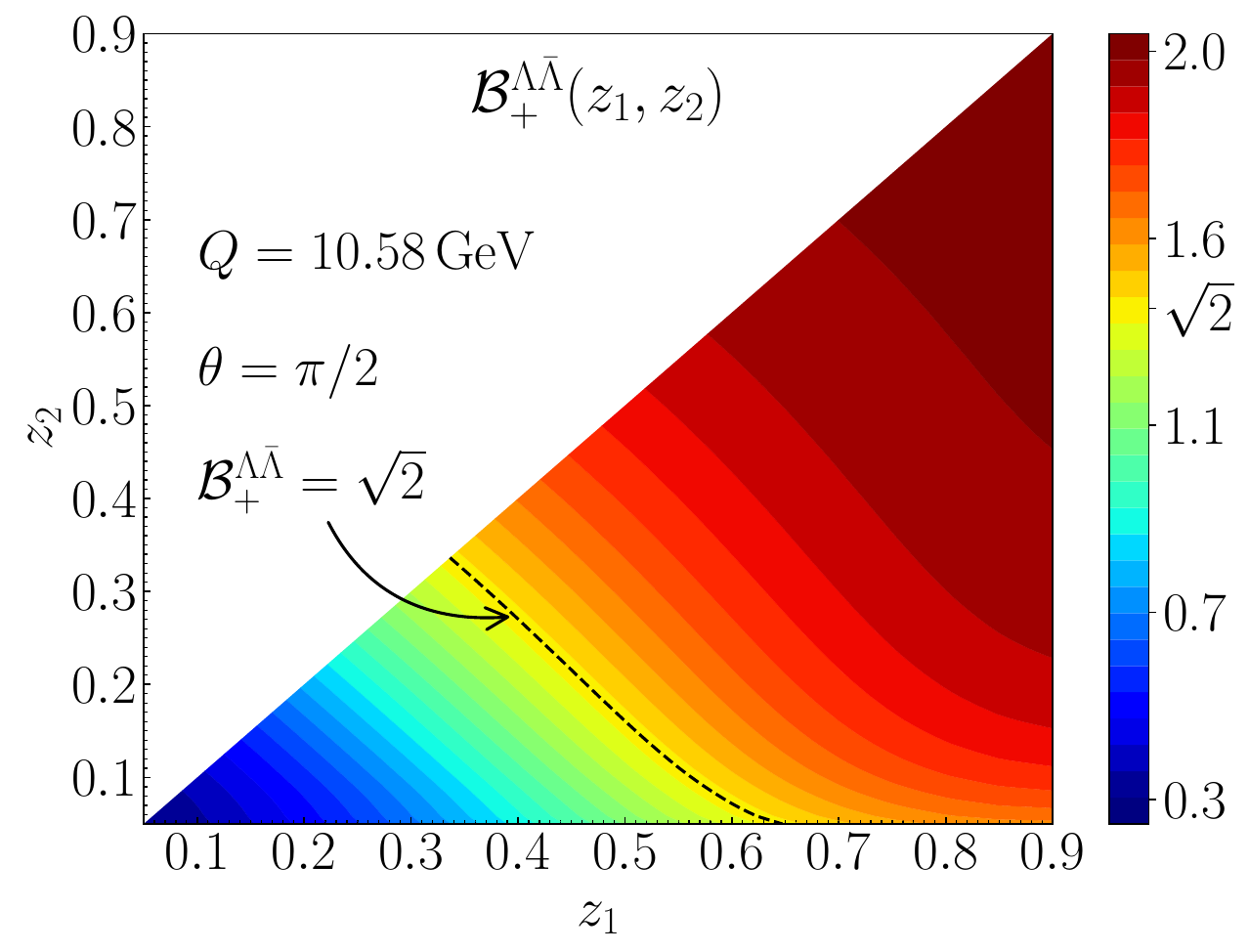}}\\[1ex]
        \vspace{0.2cm}
        \hspace{-0.5cm}
        \subfloat[three scenarios, $Q=10.58$~GeV, $\theta=\theta_0$]{\includegraphics[width=0.9\linewidth]{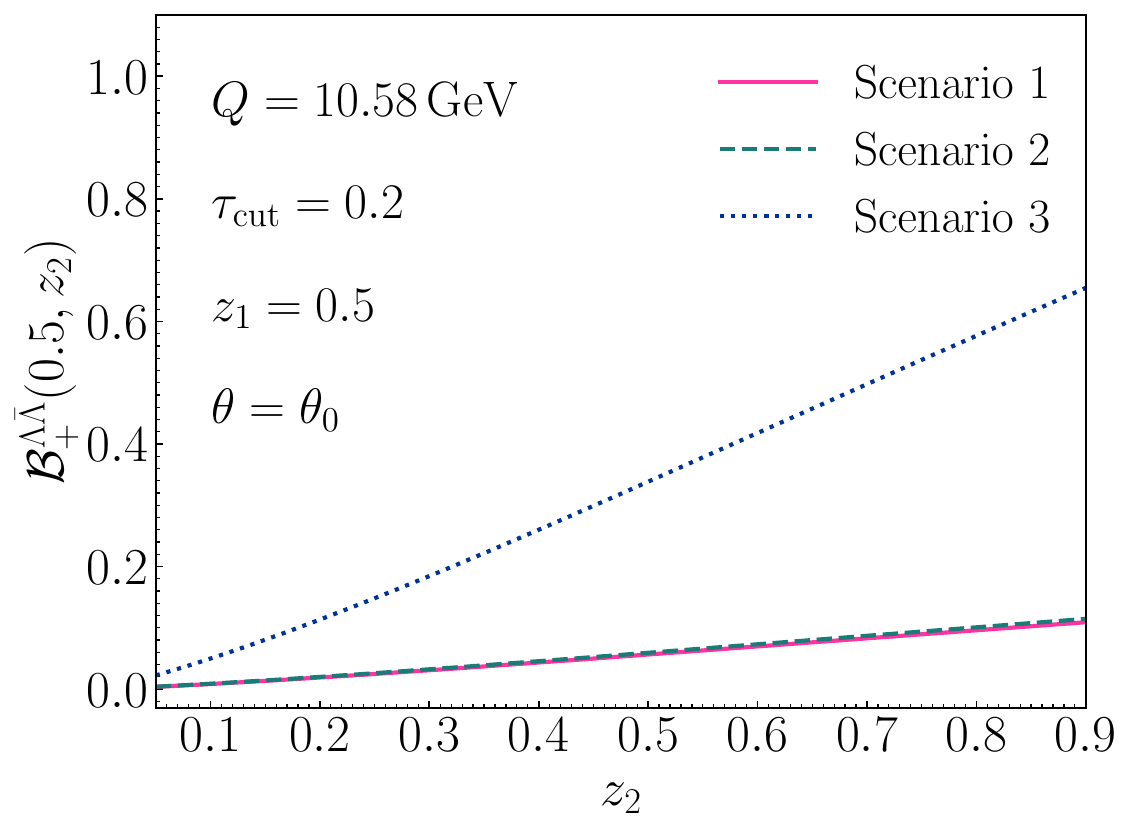}}\\[1ex]
    \end{minipage}
    \hfill
    \begin{minipage}[t]{0.48\textwidth}
        \centering
        \subfloat[positivity bound, $Q=100$~GeV, $\theta=\pi/2$]{\includegraphics[width=0.9\linewidth]{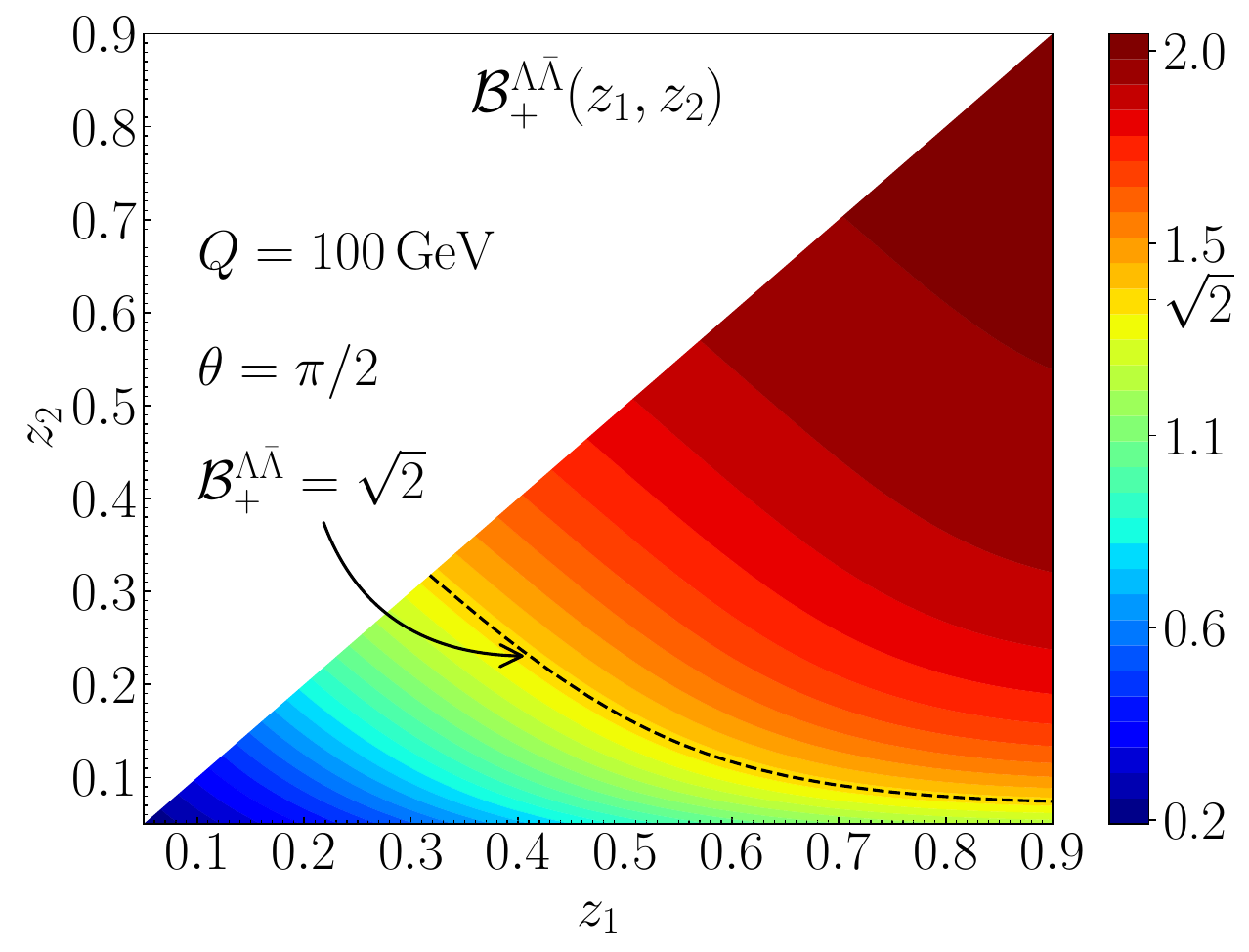}}\\[1ex]
        \vspace{0.2cm}
        \hspace{-0.5cm}
        \subfloat[three scenarios, $Q=10.58$~GeV, $\theta=\pi/2$]{\includegraphics[width=0.9\linewidth]{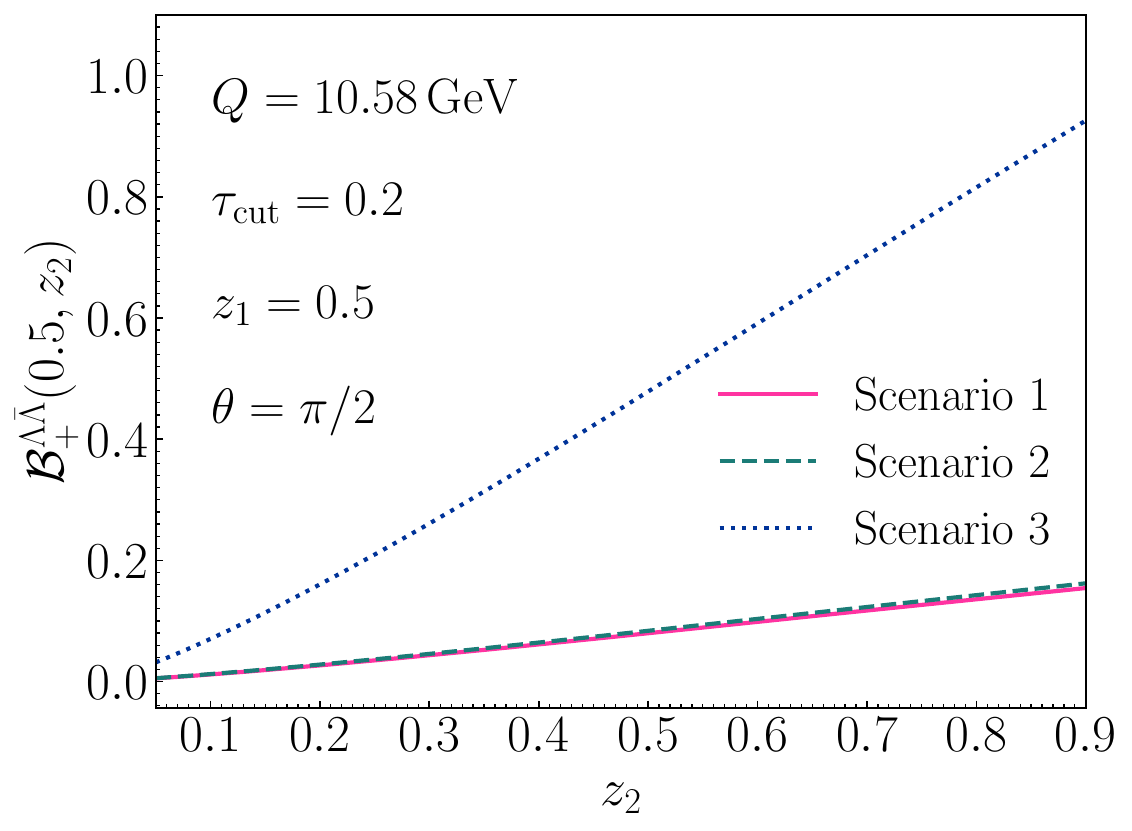}}\\[1ex]
    \end{minipage}%
    \caption{The hadronic Bell variable $\mathcal{B}^{\Lambda \bar \Lambda}_+$ as a function of $z_{1,2}$. 
    Panels (a) and (c) show contour plots of the theoretical upper limit from positivity bounds at $Q=10.58$~GeV and $100$~GeV, respectively, with $\theta=\pi/2$. The dashed line indicates the CHSH-Bell inequality violation threshold, $\mathcal{B}^{\Lambda \bar \Lambda}_+=\sqrt{2}$. Panels (b) and (d) display $\mathcal{B}_+^{\Lambda\bar{\Lambda}}$ as a function of $z_2$ at fixed $z_1=0.5$ and $Q=10.58$~GeV for the three FF scenarios. These panels compare the three FF models: Scenario 1 (magenta solid), Scenario 2 (teal dashed), and Scenario 3 (dark blue dotted). Panel (b) is evaluated at the partonic violation threshold ($\theta=\theta_0$ and $\mathcal{B}_+^{q\bar{q}}=\sqrt{2}$), while panel (d) is at the point of maximal partonic entanglement ($\theta=\pi/2$ and $\mathcal{B}_+^{q\bar{q}}=2$). }
    \label{fig:Bell_nonlocality_positivity}
\end{figure}

In this section, we evaluate the Bell variable $\mathcal{B}^{\Lambda \bar \Lambda}_+$ for the final-state $\Lambda\bar{\Lambda}$ pair using our NNLL resummed cross sections. To quantify the maximum possible spin correlation that can survive the perturbative evolution and non-perturbative hadronization process, we start from the ideal partonic baseline of a maximally entangled $q\bar{q}$ state where $\mathcal{B}^{q \bar q}_+=2$ with $\theta=\pi/2$ in Eq.~\eqref{eq:bell_parton} and use the positivity bounds on the FFs in Eq.~\eqref{eq:positivity} to establish a theoretical upper limit on the hadronic Bell variable.

Our predictions for $\mathcal{B}^{\Lambda \bar \Lambda}_+$ as a function of the momentum fractions $z_{1,2}$ are shown in figure~\ref{fig:Bell_nonlocality_positivity}. Panels (a) and (c) display the theoretical upper limit obtained from the positivity bounds on the FFs, evaluated at $\theta=\pi/2$, for $Q=10.58$~GeV and $Q=100$~GeV, respectively. The dashed line indicates the threshold for Bell inequality violation ($\mathcal{B}^{\Lambda \bar \Lambda}_+=\sqrt{2}$). We observe that even under these ideal hadronization assumptions, the Bell variable is suppressed below the partonic maximum of $2$. This demonstrates that the perturbative shower itself acts as a decohering environment. As expected, this decoherence is least severe at large $z_{1,2}$, where the hadron carries most of the parent parton's momentum and, consequently, its spin information. In this region, the system's spin correlations strengthen, enhancing the prospects for observing Bell violation. Moreover, a comparison between the results at $Q=10.58$~GeV and $Q=100$~GeV reveals a significant energy dependence. At the higher energy scale, the increased phase space for radiative emissions leads to a more pronounced decoherence effect, resulting in a stronger suppression of entanglement. This is reflected in our results by the threshold for Bell inequality violation (the dashed lines) shifting to larger momentum fractions, $z_{1,2}$. Consequently, at higher energies, sizable entanglement is expected to survive only in the kinematic regime where the final-state hadrons carry a very large fraction of the parent parton's momentum.

In panels (b) and (d), we show the predictions for $\mathcal{B}_+^{\Lambda \bar \Lambda}$ from three FF scenarios at $Q=10.58$~GeV, fixing $z_1=0.5$ and varying $z_2$. To probe the decoherence effects under different initial conditions, we evaluate the results at two key scattering angles: panel (b) corresponds to $\theta=\theta_0$ with $\theta_0=\arctan(\sqrt{2+2\sqrt{2}})$ or $\pi-\arctan\big(\sqrt{2+2\sqrt{2}}\,\big)$, which represents the threshold for partonic Bell violation ($\mathcal{B}_+^{q\bar{q}}=\sqrt{2}$). Panel (d) corresponds to $\theta=\pi/2$, where the partonic system is maximally entangled ($\mathcal{B}_+^{q\bar{q}}=2$). The three scenarios are shown as magenta solid (Scenario 1), teal dashed (Scenario 2), and dark blue dotted (Scenario 3) lines. Two key features are apparent in these model predictions. First, the results from Scenario 1 and Scenario 2 are comparable, as both models are dominated by the strange quark's transversity contribution. Second, in all three scenarios, the hadronic $\mathcal{B}_+^{\Lambda \bar \Lambda}$ remains far below the violation threshold of $\sqrt{2}$. This obvious contrast between the positivity bound calculation and these model predictions underscores the power of this observable, indicating that non-perturbative hadronization is a dominant source of decoherence. Future measurements of the kinematic dependence of $\mathcal{B}_+^{\Lambda \bar \Lambda}$ could therefore provide a unique experimental handle to disentangle perturbative and non-perturbative sources of decoherence in QCD.

\section{Summary}\label{sec:Sum}

In this work, we have presented a comprehensive theoretical analysis of spin correlations in $\Lambda\bar{\Lambda}$ production in $e^+e^-$ annihilation, providing the first state-of-the-art predictions for measurements at the Belle II experiment. Motivated by the long-standing importance of hyperon polarization in understanding QCD spin dynamics and the need for a precise theoretical framework, we have gone beyond parton model calculations by systematically incorporating complete QCD evolution effects with a thrust cut.

Our primary technical achievement is the development of a factorization theorem for the production of polarized hadron pairs in dijet events, formulated within SCET. This framework, which requires the introduction of polarized FJFs, enabled us to perform the all-orders resummation of large logarithms associated with a thrust cut to NNLL accuracy. Our numerical results for the longitudinal ($C_{LL}$) and transverse ($C_{TT}$) spin correlations demonstrate robust perturbative convergence and significantly reduced theoretical uncertainties compared to lower-order predictions, providing a solid foundation for comparison with future experimental data.

A novel and central aspect of our study is the connection between these traditional QCD observables and the formalism of quantum information theory. We established a direct mapping between the experimentally accessible spin correlation, $C_{TT}$, and a testable CHSH-Bell inequality. This powerful result reframes the measurement of spin correlations in a new light, allowing $C_{TT}$ to be interpreted as a direct, quantitative probe of the quantum decoherence induced by the fragmentation and hadronization process. Our analysis indicates that while the initial partonic state is maximally entangled, the subsequent perturbative shower and non-perturbative hadronization effects suppress the observable spin correlations, and we have quantified this suppression for various models of the fragmentation process.

Looking forward, this work opens several new avenues for research. The formalism developed here can be extended to other final-state hadrons and different collision systems, such as $ep$ and $pp$ collisions. On the phenomenological side, precise experimental measurements of $C_{LL}$ and $C_{TT}$ from Belle II and future electron-ion colliders will provide crucial data to constrain the poorly known polarized FFs and to provide crucial tests of the decoherence effects predicted here. The synergy between precision QCD calculations and concepts from quantum information science provides a rich and exciting frontier, promising deeper insights into the fundamental quantum nature of the strong interaction.

\acknowledgments

The authors thank Jiayin Gu, Wanchen Li, Lian-Tao Wang, Fang Xu, Lei Yang, and Si-Xiang Yang for helpful discussions. This work is supported by the National Science Foundations of China under Grant No. 12275052, No. 12147101, and No. 12405156, and the Innovation Program for Quantum Science and Technology under grant No. 2024ZD0300101. It is also partially supported by the Shandong Province Natural Science Foundation under grant No.~2023HWYQ-011 and No.~ZFJH202303, and the Taishan fellowship of Shandong Province for junior scientists. The authors gratefully acknowledge the valuable discussions and insights provided by the members of the China Collaboration of Precision Testing and New Physics.

\appendix

\section{Matching coefficients of FJFs}\label{app:FJFs}

In this appendix, we outline the one-loop calculation of the matching coefficients, $\mathcal{J}^{\mathcal{P}}_{iq}$. The core principle is that these perturbative coefficients can be extracted directly from the partonic FJFs, where the final-state hadron is replaced by a parton (a quark or gluon). This simplification arises because partonic FFs are trivial in dimensional regularization, i.e., $\mathcal{D}^{\mathcal{P}}_{i/j}(z, \mu) = \delta_{ij}\delta(1-z)$, as all loop corrections correspond to scaleless integrals that vanish. Consequently, the matching relation \eqref{eq:conv} reduces to a direct equality between the partonic FJF and the desired matching coefficient
\begin{align}
    \mathcal{G}^{\mathcal{P}}_{i/q}\left(z,\frac{u}{Q^2}\right)= \mathcal{J}^{\mathcal{P}}_{iq}\left(z,\frac{u}{Q^2}\right)\,.
\end{align}
Our task thus reduces to the explicit one-loop calculation of the partonic FJF, $\mathcal{G}^{\mathcal{P}}_{i/q}$. Applying the translation operator to the quark field in Eq.~\eqref{eq:FJFs} allows the partonic FJF to be expressed as
\begin{align}
    \mathcal{G}^U_{q/q}(z,s)
    &= \frac{1}{p_-} \delta(k_+-p_+-P^{X}_+) \delta(Q-p_--P^{X}_-) \nn\\
    &\quad \times \sum_X \frac{1}{2N_c} \mathrm{Tr}\left[\frac{\slashed{\bar{n}}}{2} \langle 0 | \chi_n(0)| q(p) X \rangle \langle q(p) X | \bar{\chi}_n(0) | 0 \rangle \right]\,.
\end{align}
The operator definition involves a squared matrix element that can be evaluated perturbatively. At the LO, using $s = k_+ Q$ and $p_- = z Q$, we find 
\begin{align}
    \mathcal{G}^{U(0)}_{q/q}(z,s)
    &= \frac{1}{p_-}\delta(k_+)\delta(Q-p_-) \frac{1}{2} \mathrm{Tr}\left(\frac{\slashed{\bar{n}}}{2}\slashed{p} \right) = \delta(s)\delta(1-z)\,.
\end{align}
This follows from the on-shell condition, $p^2 = 0$, combined with our frame choice which sets $p_\perp = 0$, thus requiring $p_+ = 0$. To facilitate the resummation of logarithms associated with the jet mass, we perform a Laplace transform \eqref{eq:laptransf} with respect to $s$,
\begin{align}
\mathcal{G}_{i/q}^{\mathcal{P}}\left(z,\frac{u}{Q^2}\right) = \int_0^{\infty} \md s\, \exp\left(-\frac{su}{e^{\gamma_E}Q^2}\right) \mathcal{G}_{i/q}^{\mathcal{P}}(z,s)\,.
\end{align}
In the Laplace space, the LO results are then simplified as
\begin{align}
    \mathcal{G}^{U(0)}_{q/q}\left(z, \frac{u}{Q^2} \right) = \delta(1-z)\,.
\end{align}

\begin{figure}[t]
    \centering
    \includegraphics[width=0.33\linewidth]{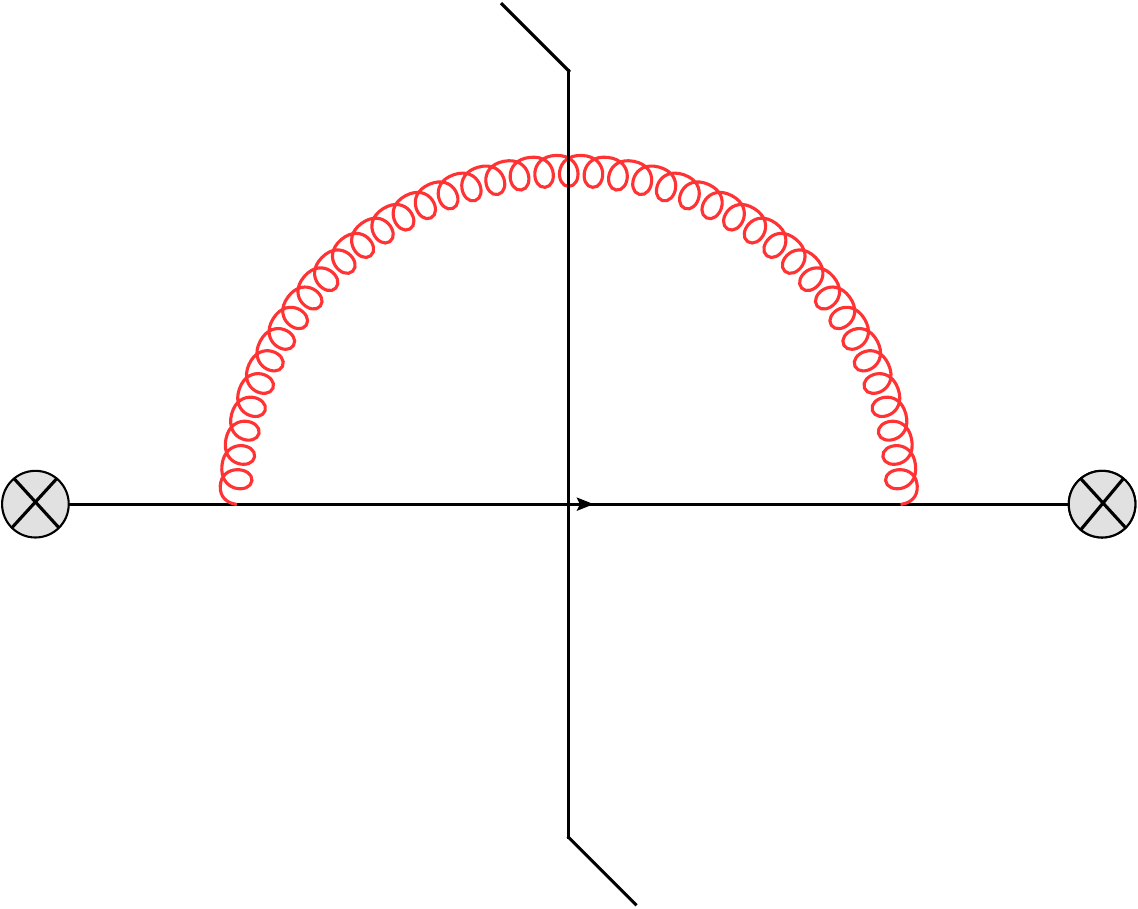}
    \caption{Feynman diagram contributing to the one-loop FJF in the light-cone gauge.}
    \label{fig:nlofjf}
\end{figure}

At the NLO, the FJFs receive contributions from real gluon emission. Denoting the gluon momentum by $p^g$, the corresponding expression is given by
\begin{align}\label{eq:appfjf}
    \mathcal{G}^{U(1)}_{q/q}(z,s)
    &= \frac{1}{p_-} \delta(k_+ - p_+ - p_+^g)\delta(p^g_- - (1-z)Q) \nn\\
    &\quad \times \frac{1}{2N_c} \mathrm{Tr}\left[\frac{\slashed{\bar{n}}}{2} \langle 0 | \chi_n(0) | q(p) g(p^g) \rangle \langle q(p) g(p^g) | \bar{\chi}_n(0) | 0 \rangle \right]\,,
\end{align}
where we replace the intermediate state $|X\rangle$ with the gluon state $|g(p^g) \rangle$. Working in light-cone gauge, where   Wilson line diagrams vanish, the calculation simplifies to the Feynman diagram shown in figure~\ref{fig:nlofjf}. The resulting expression for the one-loop partonic FJF in the Laplace space is 
\begin{align}
 \mathcal{G}^{U(1)}_{q/q}\left(z,\frac{u}{Q^2}\right)  
 &= \frac{1}{z} \int \frac{\md^{d} p^g}{(2\pi)^{d-1}} \theta(p^{g}_0)\delta(p^{g}_+p^{g}_- - |\boldsymbol{p}^{g}_T|^2) \delta(p_-^g - (1-z)Q) \nn\\
 &\quad \times g_{s,0}^2 \frac{1}{2s^2} C_F\operatorname{Tr}\left(\frac{\slashed{\bar{n}}}{2} \slashed{k} \gamma^{\mu} \slashed{p} \gamma^{\nu} \slashed{k} \right) d_{\mu\nu}(p^g, \bar{n}) \exp\left(-\frac{su}{Q^2 e^{\gamma_E}}\right)\,,
\end{align}
where $s = k^2=|\boldsymbol{p}^{g}_T|^2/(z(1-z))$, $d = 4 - 2\epsilon$ and the bare coupling $g_{s,0}^2 = 4\pi\alpha_s \left(\mu^2 e^{\gamma_E}/\\(4\pi)\right)^{\epsilon}$. Here in the light-cone gauge, we use
\begin{align}
d^{\mu\nu}(p, \bar{n}) = -g^{\mu\nu} + \frac{\bar{n}^\mu p^\nu + \bar{n}^\nu p^\mu}{\bar{n} \cdot p}.
\end{align}

For the longitudinal and transversely polarized channels ($\mathcal{P}=L,T$), the calculation proceeds similarly. The appropriate spin dependence is introduced by inserting the corresponding Dirac projectors, $\slashed{\bar{n}}\gamma_5 $ and $\slashed{\bar{n}}\gamma_5 \slashed{S}_\perp$, into the quark traces of the squared matrix element. This yields the compact expression:
\begin{align}
    \mathcal{G}^{\mathcal{P}(1)}_{i/q}(z,\frac{u}{Q^2})  
    &= \frac{1}{z} \int \frac{\md^{d} p^g}{(2\pi)^{d-1}} \theta(p^g_0)\delta(p^g_+p^g_- - |\boldsymbol{p}^g_T|^2) \delta(p^g_-/Q - (1-z)) \nn\\
    &\quad \times g_{s,0}^2 \frac{1}{s} p^{\mathcal{P}(0)}_{iq}(z,\epsilon) \exp\left(-\frac{su}{Q^2 e^{\gamma_E}}\right)\,.
\end{align}
The leading order unpolarized and polarized splitting functions are given by \cite{Vogelsang:1996im, Vogelsang:1997ak}
\begin{align}
p^{U(0)}_{qq}(z,\epsilon) &= 2C_F\left[\frac{1+z^2}{1-z} - \epsilon(1-z)\right], &
p^{U(0)}_{gq}(z,\epsilon) &= p^U_{qq}(1-z,\epsilon), \nn\\
p^{L(0)}_{qq}(z,\epsilon) &= 2C_F\left[\frac{1+z^2}{1-z} - \epsilon(1-z)\right], &
p^{L(0)}_{gq}(z,\epsilon) &= 2C_F[2 - z + 2\epsilon(1-z)], \nn\\
p^{T(0)}_{qq}(z,\epsilon) &= 2C_F\left[\frac{2z}{1-z}\right]\,.
\end{align}
After performing the phase space integrals, we arrive at the one-loop matching coefficients $\mathcal{J}^{\mathcal{P}}_{iq}$ as given in Eq.~\eqref{eq:matching_coefficients}.

\section{Resummation ingredients}\label{app:func}

This appendix collects the standard expressions required for the NNLL resummation presented in Sec.~\ref{sec:thrust}. The one-loop perturbative matching coefficients of FJFs have been given in Eq.~\eqref{eq:matching_coefficients}. The NLO hard and Laplace-transformed soft functions are given by
\begin{align}\label{eq:loop-hard}
    H\left(Q,\mu\right) = 1+\frac{\alpha_s C_F}{4\pi}\left(-2L_Q^2+6L_Q-16+\frac{7\pi^2}{3}\right)\,,
\end{align}
\begin{align}\label{eq:loop-soft}
      S_T\left(\frac{u}{Q},\mu\right) = 1+\frac{\alpha_s C_F}{4\pi}\left(-8 L_S^2-{\pi^2}\right)\,,
\end{align}
with $L_Q = \ln\left(Q^2/\mu^2\right)$ and $L_S=\ln\left(Q/u\mu \right)$. 

The resummation is governed by the anomalous dimensions, $\gamma^i(\alpha_s)$, for each component of the factorization theorem. Their perturbative expansions are given by
\begin{align}
    \gamma^i(\alpha_s) = \sum_{n=0}^\infty\gamma^i_n\left(\frac{\alpha_s}{4\pi}\right)^{n+1},\quad \text{with} \quad i = \text{cusp}, ~\mathcal{G}, ~S, ~H.
\end{align}
Here, the coefficients for the cusp and non-cusp anomalous dimensions up to NNLL accuracy are~\cite{Becher:2006mr, Korchemsky:1987wg, Moch:2004pa, Moch:2005id, Moch:2005tm, Kelley:2011ng, Hornig:2011iu, Li:2011zp}:
\begin{align}\label{eq:c1}
&\gamma_0 ^{\rm cusp}=4\,,\\
&\gamma_1^{\rm cusp}=4\left[C_{A}\left(\frac{67}{9}-\frac{\pi^{2}}{3}\right)-\frac{20}{9}T_{F}n_{f}\right]\,,\nn\\
&\gamma_2^{\rm cusp}=4 \biggr [ C_{A}^{2}\left(\frac{245}{6}-\frac{134\pi^{2}}{27}+\frac{11\pi^{4}}{45}+\frac{22\zeta_{3}}{3}\right)+C_{A}T_{F}\:n_{f}\left(-\frac{418}{27}+\frac{40\pi^{2}}{27}-\frac{56\zeta_{3}}{3}\right)\nn
\\
&\quad\quad\quad+\:T_{F}\:n_{f}\left(-\frac{55}{3}+16\zeta_{3}\right)-\frac{16}{27}\:T_{F}^{2}\:n_{f}^{2}\biggr]\,,\nn\\
&\gamma_0^\mathcal{G} = -3 C_F\,,\nn\\
&\gamma_1^\mathcal{G}=C_F^2(-\frac{3}{2}+2\pi^2-24\zeta_3)+C_FC_A\left(-\frac{1769}{54}-\frac{11\pi^2}{9}+40\zeta_3\right)+C_FT_Fn_f\left(\frac{242}{27}+\frac{4\pi^2}{9}\right),\nn\\
&\gamma_0^H=-6C_F \,,\nn\\
&\gamma_1^H=C_F^2(-3+4\pi^2-48\zeta_3)+C_FC_A\left(-\frac{961}{27}-\frac{11\pi^2}3+52\zeta_3\right)+C_FT_Fn_f\left(\frac{260}{27}+\frac{4\pi^2}3\right)\,, \notag
\end{align}
where $C_A=3\,,C_F=4/3\,,T_F=1/2\,,n_f=3\,,$ and $N_c=3$. The soft anomalous dimension can be fixed by consistency relations $\gamma^S=\gamma^H-2\gamma^\mathcal{G}$.

The solutions to the RGEs are expressed in terms of the evolution kernels $A_i(\nu,\mu)$ and $S(\nu,\mu)$, defined as
\begin{align}
 A_i(\nu,\mu)=&\frac{\gamma^i_0}{2\beta_0}\Bigg\{\ln\frac{\alpha_s(\mu)}{\alpha_s(\nu)}+\left(\frac{\gamma^i_1}{\gamma^i_0}-\frac{\beta_1}{\beta_0}\right)\frac{\alpha_s(\mu)-\alpha_s(\nu)}{4\pi} \,\nn\\
 &+\left[\frac{\gamma^i_2}{\gamma^i_0}-\frac{\beta_2}{\beta_0}-\frac{\beta_1}{\beta_0}\left(\frac{\gamma^i_1}{\gamma^i_0}-\frac{\beta_1}{\beta_0}\right)\right]\frac{\alpha_s^2(\mu)-\alpha_s^2(\nu)}{32\pi^2}\Bigg\},\nn\\
S(\nu,\mu) = & \frac{\gamma^{\text{cusp}}_0}{4\beta_0^2}\left\{\frac{4\pi}{\alpha_s(\nu)}\left(1-\frac1r-\ln r\right)+\left(\frac{\gamma^{\text{cusp}}_1}{\gamma^{\text{cusp}}_0}-\frac{\beta_1}{\beta_0}\right)(1-r+\ln r)+\frac{\beta_1}{2\beta_0}\ln^2r\right.  \,\nn\\
&+\frac{\alpha_s(\nu)}{4\pi}\Bigg[\left(\frac{\beta_1\gamma^{\text{cusp}}_1}{\beta_0\gamma^{\text{cusp}}_0}-\frac{\beta_2}{\beta_0}\right)(1-r+r\ln r)+\left(\frac{\beta_1^2}{\beta_0^2}-\frac{\beta_2}{\beta_0}\right)(1-r)\ln r \,\nn\\
&-\left.\left(\frac{\beta_1^2}{\beta_0^2}-\frac{\beta_2}{\beta_0}-\frac{\beta_1\gamma^{\text{cusp}}_1}{\beta_0\gamma^{\text{cusp}}_0}+\frac{\gamma^{\text{cusp}}_2}{\gamma^{\text{cusp}}_0}\right)\frac{(1-r)^2}2\Bigg]\right\}\,,
\end{align}
with $r \equiv \alpha_s(\mu)/\alpha_s(\nu)$.

\bibliographystyle{JHEP}
\bibliography{Refs}  

\providecommand{\href}[2]{#2}\begingroup\raggedright\begin{thebibliography}{100}

\bibitem{Bunce:1976yb}
G.~Bunce et~al., {\it {$\Lambda^0$ Hyperon Polarization in Inclusive Production by 300-GeV Protons on Beryllium.}},  {\em Phys. Rev. Lett.} {\bf 36} (1976) 1113--1116.

\bibitem{Heller:1983ia}
K.~J. Heller et~al., {\it {Polarization of $\Xi^0$ and $\Lambda$ Hyperons Produced by $400-GeV/C$ Protons}},  {\em Phys. Rev. Lett.} {\bf 51} (1983) 2025--2028.

\bibitem{Kane:1978nd}
G.~L. Kane, J.~Pumplin, and W.~Repko, {\it {Transverse Quark Polarization in Large $p_T$ Reactions, $e^+e^-$ Jets, and Leptoproduction: A Test of QCD}},  {\em Phys. Rev. Lett.} {\bf 41} (1978) 1689.

\bibitem{Liang:1997rt}
Z.-t. Liang and C.~Boros, {\it {Hyperon polarization and single spin left-right asymmetry in inclusive production processes at high-energies}},  {\em Phys. Rev. Lett.} {\bf 79} (1997) 3608--3611, [\href{http://arxiv.org/abs/hep-ph/9708488}{{\tt hep-ph/9708488}}].

\bibitem{Filippone:2001ux}
B.~W. Filippone and X.-D. Ji, {\it {The Spin structure of the nucleon}},  {\em Adv. Nucl. Phys.} {\bf 26} (2001) 1, [\href{http://arxiv.org/abs/hep-ph/0101224}{{\tt hep-ph/0101224}}].

\bibitem{DAlesio:2007bjf}
U.~D'Alesio and F.~Murgia, {\it {Azimuthal and Single Spin Asymmetries in Hard Scattering Processes}},  {\em Prog. Part. Nucl. Phys.} {\bf 61} (2008) 394--454, [\href{http://arxiv.org/abs/0712.4328}{{\tt arXiv:0712.4328}}].

\bibitem{Chen:1994ar}
K.~Chen, G.~R. Goldstein, R.~L. Jaffe, and X.-D. Ji, {\it {Probing quark fragmentation functions for spin 1/2 baryon production in unpolarized $e^+e^-$ annihilation}},  {\em Nucl. Phys. B} {\bf 445} (1995) 380--398, [\href{http://arxiv.org/abs/hep-ph/9410337}{{\tt hep-ph/9410337}}].

\bibitem{Zhang:2023ugf}
H.-C. Zhang and S.-Y. Wei, {\it {Probing the longitudinal spin transfer via dihadron polarization correlations in unpolarized $e^+e^-$ and $pp$ collisions}},  {\em Phys. Lett. B} {\bf 839} (2023) 137821, [\href{http://arxiv.org/abs/2301.04096}{{\tt arXiv:2301.04096}}].

\bibitem{Chen:2024qvx}
Z.-X. Chen, H.~Dong, and S.-Y. Wei, {\it {Dihadron helicity correlation in photon-nucleus collisions}},  {\em Phys. Rev. D} {\bf 110} (2024), no.~5 056040, [\href{http://arxiv.org/abs/2404.19202}{{\tt arXiv:2404.19202}}].

\bibitem{Yang:2024kjn}
L.~Yang, Y.-K. Song, and S.-Y. Wei, {\it {Transverse spin correlation of back-to-back dihadrons in unpolarized collisions}},  {\em Phys. Rev. D} {\bf 111} (2025), no.~5 054035, [\href{http://arxiv.org/abs/2410.20917}{{\tt arXiv:2410.20917}}].

\bibitem{Boussarie:2023izj}
R.~Boussarie et~al., {\it {TMD Handbook}},  \href{http://arxiv.org/abs/2304.03302}{{\tt arXiv:2304.03302}}.

\bibitem{Leader:2001nas}
E.~Leader, {\em {Spin in Particle Physics}}, vol.~15.
\newblock Cambridge University Press, 7, 2001.

\bibitem{STAR:2025njp}
{\bf STAR} Collaboration, {\it {Probing QCD Confinement with Spin Entanglement}},  \href{http://arxiv.org/abs/2506.05499}{{\tt arXiv:2506.05499}}.

\bibitem{Barone:2003fy}
V.~Barone and P.~G. Ratcliffe, {\em {Transverse spin physics}}.
\newblock 2003.

\bibitem{Belle-II:2018jsg}
{\bf Belle-II} Collaboration, W.~Altmannshofer et~al., {\it {The Belle II Physics Book}},  {\em PTEP} {\bf 2019} (2019), no.~12 123C01, [\href{http://arxiv.org/abs/1808.10567}{{\tt arXiv:1808.10567}}]. [Erratum: PTEP 2020, 029201 (2020)].

\bibitem{Accardi:2022oog}
A.~Accardi et~al., {\it {Opportunities for precision QCD physics in hadronization at Belle II -- a snowmass whitepaper}},  in {\em {Snowmass 2021}}, 4, 2022.
\newblock \href{http://arxiv.org/abs/2204.02280}{{\tt arXiv:2204.02280}}.

\bibitem{Belle:2008fdv}
{\bf Belle} Collaboration, R.~Seidl et~al., {\it {Measurement of Azimuthal Asymmetries in Inclusive Production of Hadron Pairs in e+e- Annihilation at s**(1/2) = 10.58-GeV}},  {\em Phys. Rev. D} {\bf 78} (2008) 032011, [\href{http://arxiv.org/abs/0805.2975}{{\tt arXiv:0805.2975}}]. [Erratum: Phys.Rev.D 86, 039905 (2012)].

\bibitem{Belle:2011cur}
{\bf Belle} Collaboration, A.~Vossen et~al., {\it {Observation of transverse polarization asymmetries of charged pion pairs in $e^+e^-$ annihilation near $\sqrt{s}=10.58$ GeV}},  {\em Phys. Rev. Lett.} {\bf 107} (2011) 072004, [\href{http://arxiv.org/abs/1104.2425}{{\tt arXiv:1104.2425}}].

\bibitem{Belle:2018ttu}
{\bf Belle} Collaboration, Y.~Guan et~al., {\it {Observation of Transverse $\Lambda/\bar{\Lambda}$ Hyperon Polarization in $e^+e^-$ Annihilation at Belle}},  {\em Phys. Rev. Lett.} {\bf 122} (2019), no.~4 042001, [\href{http://arxiv.org/abs/1808.05000}{{\tt arXiv:1808.05000}}].

\bibitem{Procura:2009vm}
M.~Procura and I.~W. Stewart, {\it {Quark Fragmentation within an Identified Jet}},  {\em Phys. Rev. D} {\bf 81} (2010) 074009, [\href{http://arxiv.org/abs/0911.4980}{{\tt arXiv:0911.4980}}]. [Erratum: Phys.Rev.D 83, 039902 (2011)].

\bibitem{Jain:2011xz}
A.~Jain, M.~Procura, and W.~J. Waalewijn, {\it {Parton Fragmentation within an Identified Jet at NNLL}},  {\em JHEP} {\bf 05} (2011) 035, [\href{http://arxiv.org/abs/1101.4953}{{\tt arXiv:1101.4953}}].

\bibitem{Procura:2014cba}
M.~Procura, W.~J. Waalewijn, and L.~Zeune, {\it {Resummation of Double-Differential Cross Sections and Fully-Unintegrated Parton Distribution Functions}},  {\em JHEP} {\bf 02} (2015) 117, [\href{http://arxiv.org/abs/1410.6483}{{\tt arXiv:1410.6483}}].

\bibitem{Lustermans:2019plv}
G.~Lustermans, J.~K.~L. Michel, F.~J. Tackmann, and W.~J. Waalewijn, {\it {Joint two-dimensional resummation in $q_{T}$ and $0$-jettiness at NNLL}},  {\em JHEP} {\bf 03} (2019) 124, [\href{http://arxiv.org/abs/1901.03331}{{\tt arXiv:1901.03331}}].

\bibitem{Kang:2020yqw}
Z.-B. Kang, D.~Y. Shao, and F.~Zhao, {\it {QCD resummation on single hadron transverse momentum distribution with the thrust axis}},  {\em JHEP} {\bf 12} (2020) 127, [\href{http://arxiv.org/abs/2007.14425}{{\tt arXiv:2007.14425}}].

\bibitem{Makris:2020ltr}
Y.~Makris, F.~Ringer, and W.~J. Waalewijn, {\it {Joint thrust and TMD resummation in electron-positron and electron-proton collisions}},  {\em JHEP} {\bf 02} (2021) 070, [\href{http://arxiv.org/abs/2009.11871}{{\tt arXiv:2009.11871}}].

\bibitem{Boglione:2020auc}
M.~Boglione and A.~Simonelli, {\it {Factorization of $e^+e^- \to H\,X$ cross section, differential in $z_h$, $P_T$ and thrust, in the $2$-jet limit}},  {\em JHEP} {\bf 02} (2021) 076, [\href{http://arxiv.org/abs/2011.07366}{{\tt arXiv:2011.07366}}].

\bibitem{Gamberg:2021iat}
L.~Gamberg, Z.-B. Kang, D.~Y. Shao, J.~Terry, and F.~Zhao, {\it {Transverse $\Lambda$ polarization in $e^+e^-$ collisions}},  {\em Phys. Lett. B} {\bf 818} (2021) 136371, [\href{http://arxiv.org/abs/2102.05553}{{\tt arXiv:2102.05553}}].

\bibitem{Boglione:2021wov}
M.~Boglione and A.~Simonelli, {\it {Kinematic regions in the $e^+e^- \to h \, X$ factorized cross section in a $2$-jet topology with thrust}},  {\em JHEP} {\bf 02} (2022) [\href{http://arxiv.org/abs/2109.11497}{{\tt arXiv:2109.11497}}].

\bibitem{Boglione:2023duo}
M.~Boglione and A.~Simonelli, {\it {Full treatment of the thrust distribution in single inclusive $e^+e^- \to h \, X$ processes}},  {\em JHEP} {\bf 09} (2023) 006, [\href{http://arxiv.org/abs/2306.02937}{{\tt arXiv:2306.02937}}].

\bibitem{Fang:2025dee}
S.~Fang, S.~Lin, D.~Y. Shao, and J.~Zhou, {\it {Nucleon Tomography with 0-jettiness}},  \href{http://arxiv.org/abs/2506.15962}{{\tt arXiv:2506.15962}}.

\bibitem{Bauer:2000yr}
C.~W. Bauer, S.~Fleming, D.~Pirjol, and I.~W. Stewart, {\it {An Effective field theory for collinear and soft gluons: Heavy to light decays}},  {\em Phys. Rev. D} {\bf 63} (2001) 114020, [\href{http://arxiv.org/abs/hep-ph/0011336}{{\tt hep-ph/0011336}}].

\bibitem{Bauer:2001ct}
C.~W. Bauer and I.~W. Stewart, {\it {Invariant operators in collinear effective theory}},  {\em Phys. Lett. B} {\bf 516} (2001) 134--142, [\href{http://arxiv.org/abs/hep-ph/0107001}{{\tt hep-ph/0107001}}].

\bibitem{Bauer:2001yt}
C.~W. Bauer, D.~Pirjol, and I.~W. Stewart, {\it {Soft collinear factorization in effective field theory}},  {\em Phys. Rev. D} {\bf 65} (2002) 054022, [\href{http://arxiv.org/abs/hep-ph/0109045}{{\tt hep-ph/0109045}}].

\bibitem{Bauer:2002nz}
C.~W. Bauer, S.~Fleming, D.~Pirjol, I.~Z. Rothstein, and I.~W. Stewart, {\it {Hard scattering factorization from effective field theory}},  {\em Phys. Rev. D} {\bf 66} (2002) 014017, [\href{http://arxiv.org/abs/hep-ph/0202088}{{\tt hep-ph/0202088}}].

\bibitem{Beneke:2002ph}
M.~Beneke, A.~P. Chapovsky, M.~Diehl, and T.~Feldmann, {\it {Soft collinear effective theory and heavy to light currents beyond leading power}},  {\em Nucl. Phys. B} {\bf 643} (2002) 431--476, [\href{http://arxiv.org/abs/hep-ph/0206152}{{\tt hep-ph/0206152}}].

\bibitem{Abel:1992kz}
S.~A. Abel, M.~Dittmar, and H.~K. Dreiner, {\it {Testing locality at colliders via Bell's inequality?}},  {\em Phys. Lett. B} {\bf 280} (1992) 304--312.

\bibitem{ATLAS:2023fsd}
{\bf ATLAS} Collaboration, G.~Aad et~al., {\it {Observation of quantum entanglement with top quarks at the ATLAS detector}},  {\em Nature} {\bf 633} (2024), no.~8030 542--547, [\href{http://arxiv.org/abs/2311.07288}{{\tt arXiv:2311.07288}}].

\bibitem{CMS:2024pts}
{\bf CMS} Collaboration, A.~Hayrapetyan et~al., {\it {Observation of quantum entanglement in top quark pair production in proton\textendash{}proton collisions at $\sqrt{s} = 13$ TeV}},  {\em Rept. Prog. Phys.} {\bf 87} (2024), no.~11 117801, [\href{http://arxiv.org/abs/2406.03976}{{\tt arXiv:2406.03976}}].

\bibitem{Severi:2021cnj}
C.~Severi, C.~D.~E. Boschi, F.~Maltoni, and M.~Sioli, {\it {Quantum tops at the LHC: from entanglement to Bell inequalities}},  {\em Eur. Phys. J. C} {\bf 82} (2022), no.~4 285, [\href{http://arxiv.org/abs/2110.10112}{{\tt arXiv:2110.10112}}].

\bibitem{Altakach:2022ywa}
M.~M. Altakach, P.~Lamba, F.~Maltoni, K.~Mawatari, and K.~Sakurai, {\it {Quantum information and CP measurement in $H\to {\tau}^+{\tau}^-$ at future lepton colliders}},  {\em Phys. Rev. D} {\bf 107} (2023), no.~9 093002, [\href{http://arxiv.org/abs/2211.10513}{{\tt arXiv:2211.10513}}].

\bibitem{Aguilar-Saavedra:2022uye}
J.~A. Aguilar-Saavedra and J.~A. Casas, {\it {Improved tests of entanglement and Bell inequalities with LHC tops}},  {\em Eur. Phys. J. C} {\bf 82} (2022), no.~8 666, [\href{http://arxiv.org/abs/2205.00542}{{\tt arXiv:2205.00542}}].

\bibitem{Ehataht:2023zzt}
K.~Ehat{\"a}ht, M.~Fabbrichesi, L.~Marzola, and C.~Veelken, {\it {Probing entanglement and testing Bell inequality violation with $e^+e^- \to \tau^+\tau^-$ at Belle II}},  {\em Phys. Rev. D} {\bf 109} (2024), no.~3 032005, [\href{http://arxiv.org/abs/2311.17555}{{\tt arXiv:2311.17555}}].

\bibitem{Fabbrichesi:2024wcd}
M.~Fabbrichesi and L.~Marzola, {\it {Quantum tomography with $\tau$ leptons at the FCC-ee: Entanglement, Bell inequality violation, $\sin\theta_W$, and anomalous couplings}},  {\em Phys. Rev. D} {\bf 110} (2024), no.~7 076004, [\href{http://arxiv.org/abs/2405.09201}{{\tt arXiv:2405.09201}}].

\bibitem{Barr:2022wyq}
A.~J. Barr, P.~Caban, and J.~Rembieli\'nski, {\it {Bell-type inequalities for systems of relativistic vector bosons}},  {\em Quantum} {\bf 7} (2023) 1070, [\href{http://arxiv.org/abs/2204.11063}{{\tt arXiv:2204.11063}}].

\bibitem{Ashby-Pickering:2022umy}
R.~Ashby-Pickering, A.~J. Barr, and A.~Wierzchucka, {\it {Quantum state tomography, entanglement detection and Bell violation prospects in weak decays of massive particles}},  {\em JHEP} {\bf 05} (2023) 020, [\href{http://arxiv.org/abs/2209.13990}{{\tt arXiv:2209.13990}}].

\bibitem{Aguilar-Saavedra:2022wam}
J.~A. Aguilar-Saavedra, A.~Bernal, J.~A. Casas, and J.~M. Moreno, {\it {Testing entanglement and Bell inequalities in $H \to ZZ$}},  {\em Phys. Rev. D} {\bf 107} (2023), no.~1 016012, [\href{http://arxiv.org/abs/2209.13441}{{\tt arXiv:2209.13441}}].

\bibitem{Fabbrichesi:2023cev}
M.~Fabbrichesi, R.~Floreanini, E.~Gabrielli, and L.~Marzola, {\it {Bell inequalities and quantum entanglement in weak gauge boson production at the LHC and future colliders}},  {\em Eur. Phys. J. C} {\bf 83} (2023), no.~9 823, [\href{http://arxiv.org/abs/2302.00683}{{\tt arXiv:2302.00683}}].

\bibitem{Du:2024sly}
Y.~Du, X.-G. He, C.-W. Liu, and J.-P. Ma, {\it {Impact of parity violation on quantum entanglement and Bell nonlocality}},  \href{http://arxiv.org/abs/2409.15418}{{\tt arXiv:2409.15418}}.

\bibitem{Fabbri:2023ncz}
F.~Fabbri, J.~Howarth, and T.~Maurin, {\it {Isolating semi-leptonic $H\rightarrow WW^{*}$decays for Bell inequality tests}},  {\em Eur. Phys. J. C} {\bf 84} (2024), no.~1 20, [\href{http://arxiv.org/abs/2307.13783}{{\tt arXiv:2307.13783}}].

\bibitem{Han:2023fci}
T.~Han, M.~Low, and T.~A. Wu, {\it {Quantum entanglement and Bell inequality violation in semi-leptonic top decays}},  {\em JHEP} {\bf 07} (2024) 192, [\href{http://arxiv.org/abs/2310.17696}{{\tt arXiv:2310.17696}}].

\bibitem{Bi:2023uop}
Q.~Bi, Q.-H. Cao, K.~Cheng, and H.~Zhang, {\it {New observables for testing Bell inequalities in W boson pair production}},  {\em Phys. Rev. D} {\bf 109} (2024), no.~3 036022, [\href{http://arxiv.org/abs/2307.14895}{{\tt arXiv:2307.14895}}].

\bibitem{Wu:2024asu}
S.~Wu, C.~Qian, Q.~Wang, and X.-R. Zhou, {\it {Bell nonlocality and entanglement in $e^+e^- \to Y\bar{Y}$ at BESIII}},  {\em Phys. Rev. D} {\bf 110} (2024), no.~5 054012, [\href{http://arxiv.org/abs/2406.16298}{{\tt arXiv:2406.16298}}].

\bibitem{Wu:2024mtj}
S.~Wu, C.~Qian, Y.-G. Yang, and Q.~Wang, {\it {Generalized Quantum Measurement in Spin-Correlated Hyperon-Antihyperon Decays}},  {\em Chin. Phys. Lett.} {\bf 41} (2024), no.~11 110301, [\href{http://arxiv.org/abs/2402.16574}{{\tt arXiv:2402.16574}}].

\bibitem{Pei:2025non}
J.~Pei, Y.~Fang, L.~Wu, D.~Xu, M.~Biyabi, and T.~Li, {\it {Quantum Entanglement Theory and Its Generic Searches in High Energy Physics}},  \href{http://arxiv.org/abs/2505.09280}{{\tt arXiv:2505.09280}}.

\bibitem{Afik:2025grr}
Y.~Afik, Y.~Kats, J.~R.~M. de~Nova, A.~Soffer, and D.~Uzan, {\it {Entanglement and Bell nonlocality with bottom-quark pairs at hadron colliders}},  {\em Phys. Rev. D} {\bf 111} (2025), no.~11 L111902, [\href{http://arxiv.org/abs/2406.04402}{{\tt arXiv:2406.04402}}].

\bibitem{Ruzi:2025jql}
A.~Ruzi, Y.~Wu, R.~Ding, and Q.~Li, {\it {Searching Quantum Entanglement in $pp\to ZZ$ process}},  \href{http://arxiv.org/abs/2506.16077}{{\tt arXiv:2506.16077}}.

\bibitem{Goncalves:2025xer}
D.~Gon{\c{c}}alves, A.~Kaladharan, and A.~Navarro, {\it {Higher-Order Corrections to Quantum Observables in $h\to WW^*$}},  \href{http://arxiv.org/abs/2506.19951}{{\tt arXiv:2506.19951}}.

\bibitem{Cheng:2025zcf}
K.~Cheng, T.~Han, M.~Low, and T.~A. Wu, {\it {Quantum Tomography in Neutral Meson and Antimeson Systems}},  \href{http://arxiv.org/abs/2507.12513}{{\tt arXiv:2507.12513}}.

\bibitem{Morales:2023gow}
R.~A. Morales, {\it {Exploring Bell inequalities and quantum entanglement in vector boson scattering}},  {\em Eur. Phys. J. Plus} {\bf 138} (2023), no.~12 1157, [\href{http://arxiv.org/abs/2306.17247}{{\tt arXiv:2306.17247}}].

\bibitem{Morales:2024jhj}
R.~A. Morales, {\it {Tripartite entanglement and Bell non-locality in loop-induced Higgs boson decays}},  {\em Eur. Phys. J. C} {\bf 84} (2024), no.~6 581, [\href{http://arxiv.org/abs/2403.18023}{{\tt arXiv:2403.18023}}].

\bibitem{Goncalves:2025qem}
D.~Gon{\c{c}}alves, A.~Kaladharan, F.~Krauss, and A.~Navarro, {\it {Quantum Entanglement is Quantum: ZZ Production at the LHC}},  \href{http://arxiv.org/abs/2505.12125}{{\tt arXiv:2505.12125}}.

\bibitem{Dong:2023xiw}
Z.~Dong, D.~Gon{\c{c}}alves, K.~Kong, and A.~Navarro, {\it {Entanglement and Bell inequalities with boosted tt{\textasciimacron}}},  {\em Phys. Rev. D} {\bf 109} (2024), no.~11 115023, [\href{http://arxiv.org/abs/2305.07075}{{\tt arXiv:2305.07075}}].

\bibitem{Guo:2024jch}
Y.~Guo, X.~Liu, F.~Yuan, and H.~X. Zhu, {\it {Long-Range Azimuthal Correlation, Entanglement, and Bell Inequality Violation by Spinning Gluons at the Large Hadron Collider}},  {\em Research} {\bf 2025} (2025) 0552, [\href{http://arxiv.org/abs/2406.05880}{{\tt arXiv:2406.05880}}].

\bibitem{Cheng:2025cuv}
K.~Cheng and B.~Yan, {\it {Bell Inequality Violation of Light Quarks in Dihadron Pair Production at Lepton Colliders}},  {\em Phys. Rev. Lett.} {\bf 135} (2025), no.~1 011902, [\href{http://arxiv.org/abs/2501.03321}{{\tt arXiv:2501.03321}}].

\bibitem{vonKuk:2025kbv}
R.~von Kuk, K.~Lee, J.~K.~L. Michel, and Z.~Sun, {\it {Towards a Quantum Information Theory of Hadronization: Dihadron Fragmentation and Neutral Polarization in Heavy Baryons}},  \href{http://arxiv.org/abs/2503.22607}{{\tt arXiv:2503.22607}}.

\bibitem{Qi:2025onf}
W.~Qi, Z.~Guo, and B.-W. Xiao, {\it {Studying Maximal Entanglement and Bell Nonlocality at an Electron-Ion Collider}},  \href{http://arxiv.org/abs/2506.12889}{{\tt arXiv:2506.12889}}.

\bibitem{Bertlmann:2004yg}
R.~A. Bertlmann, {\it {Entanglement, Bell inequalities and decoherence in particle physics}},  {\em Lect. Notes Phys.} {\bf 689} (2006) 1--45, [\href{http://arxiv.org/abs/quant-ph/0410028}{{\tt quant-ph/0410028}}].

\bibitem{Carney:2017jut}
D.~Carney, L.~Chaurette, D.~Neuenfeld, and G.~W. Semenoff, {\it {Infrared quantum information}},  {\em Phys. Rev. Lett.} {\bf 119} (2017), no.~18 180502, [\href{http://arxiv.org/abs/1706.03782}{{\tt arXiv:1706.03782}}].

\bibitem{Carney:2017oxp}
D.~Carney, L.~Chaurette, D.~Neuenfeld, and G.~W. Semenoff, {\it {Dressed infrared quantum information}},  {\em Phys. Rev. D} {\bf 97} (2018), no.~2 025007, [\href{http://arxiv.org/abs/1710.02531}{{\tt arXiv:1710.02531}}].

\bibitem{Carney:2018ygh}
D.~Carney, L.~Chaurette, D.~Neuenfeld, and G.~Semenoff, {\it {On the need for soft dressing}},  {\em JHEP} {\bf 09} (2018) 121, [\href{http://arxiv.org/abs/1803.02370}{{\tt arXiv:1803.02370}}].

\bibitem{Neuenfeld:2018fdw}
D.~Neuenfeld, {\it {Infrared-safe scattering without photon vacuum transitions and time-dependent decoherence}},  {\em JHEP} {\bf 11} (2021) 189, [\href{http://arxiv.org/abs/1810.11477}{{\tt arXiv:1810.11477}}].

\bibitem{Semenoff:2019dqe}
G.~W. Semenoff, {\it {Entanglement and the Infrared}},  {\em Springer Proc. Math. Stat.} {\bf 335} (2019) 151--166, [\href{http://arxiv.org/abs/1912.03187}{{\tt arXiv:1912.03187}}].

\bibitem{Schlosshauer:2019ewh}
M.~Schlosshauer, {\it {Quantum decoherence}},  {\em Phys. Rept.} {\bf 831} (2019) 1--57, [\href{http://arxiv.org/abs/1911.06282}{{\tt arXiv:1911.06282}}].

\bibitem{Burgess:2024heo}
C.~P. Burgess, T.~Colas, R.~Holman, and G.~Kaplanek, {\it {Does decoherence violate decoupling?}},  {\em JHEP} {\bf 02} (2025) 204, [\href{http://arxiv.org/abs/2411.09000}{{\tt arXiv:2411.09000}}].

\bibitem{Salcedo:2024smn}
S.~A. Salcedo, T.~Colas, and E.~Pajer, {\it {The open effective field theory of inflation}},  {\em JHEP} {\bf 10} (2024) 248, [\href{http://arxiv.org/abs/2404.15416}{{\tt arXiv:2404.15416}}].

\bibitem{Salcedo:2024nex}
S.~A. Salcedo, T.~Colas, and E.~Pajer, {\it {An Open Effective Field Theory for light in a medium}},  {\em JHEP} {\bf 03} (2025) 138, [\href{http://arxiv.org/abs/2412.12299}{{\tt arXiv:2412.12299}}].

\bibitem{Aoude:2025ovu}
R.~Aoude, A.~J. Barr, F.~Maltoni, and L.~Satrioni, {\it {Decoherence effects in entangled fermion pairs at colliders}},  \href{http://arxiv.org/abs/2504.07030}{{\tt arXiv:2504.07030}}.

\bibitem{Li:2024luk}
S.~Li, W.~Shen, and J.~M. Yang, {\it {Can Bell inequalities be tested via scattering cross-section at colliders ?}},  {\em Eur. Phys. J. C} {\bf 84} (2024), no.~11 1195, [\href{http://arxiv.org/abs/2401.01162}{{\tt arXiv:2401.01162}}].

\bibitem{Bechtle:2025ugc}
P.~Bechtle, C.~Breuning, H.~K. Dreiner, and C.~Duhr, {\it {A critical appraisal of tests of locality and of entanglement versus non-entanglement at colliders}},  \href{http://arxiv.org/abs/2507.15947}{{\tt arXiv:2507.15947}}.

\bibitem{Abel:2025skj}
S.~A. Abel, H.~K. Dreiner, R.~Sengupta, and L.~Ubaldi, {\it {Colliders are Testing neither Locality via Bell's Inequality nor Entanglement versus Non-Entanglement}},  \href{http://arxiv.org/abs/2507.15949}{{\tt arXiv:2507.15949}}.

\bibitem{Low:2025aqq}
M.~Low, {\it {Addressing Local Realism through Bell Tests at Colliders}},  \href{http://arxiv.org/abs/2508.10979}{{\tt arXiv:2508.10979}}.

\bibitem{Jacob:1959at}
M.~Jacob and G.~C. Wick, {\it {On the General Theory of Collisions for Particles with Spin}},  {\em Annals Phys.} {\bf 7} (1959) 404--428.

\bibitem{DAlesio:2021dcx}
U.~D'Alesio, F.~Murgia, and M.~Zaccheddu, {\it {General helicity formalism for two-hadron production in $e^+e^-$ annihilation within a TMD approach}},  {\em JHEP} {\bf 10} (2021) 078, [\href{http://arxiv.org/abs/2108.05632}{{\tt arXiv:2108.05632}}].

\bibitem{Anselmino:2005sh}
M.~Anselmino, M.~Boglione, U.~D'Alesio, E.~Leader, S.~Melis, and F.~Murgia, {\it {The general partonic structure for hadronic spin asymmetries}},  {\em Phys. Rev. D} {\bf 73} (2006) 014020, [\href{http://arxiv.org/abs/hep-ph/0509035}{{\tt hep-ph/0509035}}].

\bibitem{Batozskaya:2023rek}
V.~Batozskaya, A.~Kupsc, N.~Salone, and J.~Wiechnik, {\it {Semileptonic decays of spin-entangled baryon-antibaryon pairs}},  {\em Phys. Rev. D} {\bf 108} (2023), no.~1 016011, [\href{http://arxiv.org/abs/2302.07665}{{\tt arXiv:2302.07665}}].

\bibitem{Perotti:2018wxm}
E.~Perotti, G.~F{\"a}ldt, A.~Kupsc, S.~Leupold, and J.~J. Song, {\it {Polarization observables in $e^+e^-$ annihilation to a baryon-antibaryon pair}},  {\em Phys. Rev. D} {\bf 99} (2019), no.~5 056008, [\href{http://arxiv.org/abs/1809.04038}{{\tt arXiv:1809.04038}}].

\bibitem{Boer:1997mf}
D.~Boer, R.~Jakob, and P.~J. Mulders, {\it {Asymmetries in polarized hadron production in $e^+ e^-$ annihilation up to order $1/Q$}},  {\em Nucl. Phys. B} {\bf 504} (1997) 345--380, [\href{http://arxiv.org/abs/hep-ph/9702281}{{\tt hep-ph/9702281}}].

\bibitem{Jaffe:1991ra}
R.~L. Jaffe and X.-D. Ji, {\it {Chiral odd parton distributions and Drell-Yan processes}},  {\em Nucl. Phys. B} {\bf 375} (1992) 527--560.

\bibitem{Kang:2023elg}
Z.-B. Kang, H.~Xing, F.~Zhao, and Y.~Zhou, {\it {Polarized fragmenting jet functions in inclusive and exclusive jet production}},  {\em JHEP} {\bf 03} (2024) 142, [\href{http://arxiv.org/abs/2311.00672}{{\tt arXiv:2311.00672}}].

\bibitem{Collins:1992kk}
J.~C. Collins, {\it {Fragmentation of transversely polarized quarks probed in transverse momentum distributions}},  {\em Nucl. Phys. B} {\bf 396} (1993) 161--182, [\href{http://arxiv.org/abs/hep-ph/9208213}{{\tt hep-ph/9208213}}].

\bibitem{Becher:2014oda}
T.~Becher, A.~Broggio, and A.~Ferroglia, {\em {Introduction to Soft-Collinear Effective Theory}}, vol.~896.
\newblock Springer, 2015.

\bibitem{Brandenburg:1996df}
A.~Brandenburg, {\it {Spin spin correlations of top quark pairs at hadron colliders}},  {\em Phys. Lett. B} {\bf 388} (1996) 626--632, [\href{http://arxiv.org/abs/hep-ph/9603333}{{\tt hep-ph/9603333}}].

\bibitem{Afik:2022kwm}
Y.~Afik and J.~R.~M. de~Nova, {\it {Quantum information with top quarks in QCD}},  {\em Quantum} {\bf 6} (2022) 820, [\href{http://arxiv.org/abs/2203.05582}{{\tt arXiv:2203.05582}}].

\bibitem{Jaffe:1996zw}
R.~L. Jaffe, {\it {Spin, twist and hadron structure in deep inelastic processes}},  in {\em {Ettore Majorana International School of Nucleon Structure: 1st Course: The Spin Structure of the Nucleon}}, pp.~42--129, 1, 1996.
\newblock \href{http://arxiv.org/abs/hep-ph/9602236}{{\tt hep-ph/9602236}}.

\bibitem{DAlesio:2022wbc}
U.~D'Alesio, F.~Murgia, and M.~Zaccheddu, {\it {General Helicity Formalism for Two-hadron Production in $e^+e^-$ Collisions and the $\Lambda$ Polarizing Fragmentation Function}},  {\em JPS Conf. Proc.} {\bf 37} (2022) 020112, [\href{http://arxiv.org/abs/2202.10056}{{\tt arXiv:2202.10056}}].

\bibitem{Anedda:2025bts}
S.~Anedda, F.~Murgia, and C.~Pisano, {\it {First insight into transverse-momentum-dependent fragmentation physics at photon-photon colliders}},  {\em Phys. Rev. D} {\bf 112} (2025), no.~1 014013, [\href{http://arxiv.org/abs/2504.12802}{{\tt arXiv:2504.12802}}].

\bibitem{Jaffe:1996wp}
R.~L. Jaffe, {\it {Polarized $\Lambda$'s in the current fragmentation region}},  {\em Phys. Rev. D} {\bf 54} (1996), no.~11 R6581--R6585, [\href{http://arxiv.org/abs/hep-ph/9605456}{{\tt hep-ph/9605456}}].

\bibitem{Farhi:1977sg}
E.~Farhi, {\it {A QCD Test for Jets}},  {\em Phys. Rev. Lett.} {\bf 39} (1977) 1587--1588.

\bibitem{Becher:2008cf}
T.~Becher and M.~D. Schwartz, {\it {A precise determination of $\alpha_s$ from LEP thrust data using effective field theory}},  {\em JHEP} {\bf 07} (2008) 034, [\href{http://arxiv.org/abs/0803.0342}{{\tt arXiv:0803.0342}}].

\bibitem{Schwartz:2014sze}
M.~D. Schwartz, {\em {Quantum Field Theory and the Standard Model}}.
\newblock Cambridge University Press, 3, 2014.

\bibitem{Kang:2016ehg}
Z.-B. Kang, F.~Ringer, and I.~Vitev, {\it {Jet substructure using semi-inclusive jet functions in SCET}},  {\em JHEP} {\bf 11} (2016) 155, [\href{http://arxiv.org/abs/1606.07063}{{\tt arXiv:1606.07063}}].

\bibitem{Kang:2016mcy}
Z.-B. Kang, F.~Ringer, and I.~Vitev, {\it {The semi-inclusive jet function in SCET and small radius resummation for inclusive jet production}},  {\em JHEP} {\bf 10} (2016) 125, [\href{http://arxiv.org/abs/1606.06732}{{\tt arXiv:1606.06732}}].

\bibitem{Jain:2011iu}
A.~Jain, M.~Procura, and W.~J. Waalewijn, {\it {Fully-Unintegrated Parton Distribution and Fragmentation Functions at Perturbative $k_T$}},  {\em JHEP} {\bf 04} (2012) 132, [\href{http://arxiv.org/abs/1110.0839}{{\tt arXiv:1110.0839}}].

\bibitem{Catani:1992ua}
S.~Catani, L.~Trentadue, G.~Turnock, and B.~R. Webber, {\it {Resummation of large logarithms in $e^+e^-$ event shape distributions}},  {\em Nucl. Phys. B} {\bf 407} (1993) 3--42.

\bibitem{Korchemsky:1999kt}
G.~P. Korchemsky and G.~F. Sterman, {\it {Power corrections to event shapes and factorization}},  {\em Nucl. Phys. B} {\bf 555} (1999) 335--351, [\href{http://arxiv.org/abs/hep-ph/9902341}{{\tt hep-ph/9902341}}].

\bibitem{Fleming:2007qr}
S.~Fleming, A.~H. Hoang, S.~Mantry, and I.~W. Stewart, {\it {Jets from massive unstable particles: Top-mass determination}},  {\em Phys. Rev. D} {\bf 77} (2008) 074010, [\href{http://arxiv.org/abs/hep-ph/0703207}{{\tt hep-ph/0703207}}].

\bibitem{Schwartz:2007ib}
M.~D. Schwartz, {\it {Resummation and NLO matching of event shapes with effective field theory}},  {\em Phys. Rev. D} {\bf 77} (2008) 014026, [\href{http://arxiv.org/abs/0709.2709}{{\tt arXiv:0709.2709}}].

\bibitem{Becher:2006nr}
T.~Becher and M.~Neubert, {\it {Threshold resummation in momentum space from effective field theory}},  {\em Phys. Rev. Lett.} {\bf 97} (2006) 082001, [\href{http://arxiv.org/abs/hep-ph/0605050}{{\tt hep-ph/0605050}}].

\bibitem{Becher:2006mr}
T.~Becher, M.~Neubert, and B.~D. Pecjak, {\it {Factorization and Momentum-Space Resummation in Deep-Inelastic Scattering}},  {\em JHEP} {\bf 01} (2007) 076, [\href{http://arxiv.org/abs/hep-ph/0607228}{{\tt hep-ph/0607228}}].

\bibitem{Sterman:1986aj}
G.~F. Sterman, {\it {Summation of Large Corrections to Short Distance Hadronic Cross-Sections}},  {\em Nucl. Phys. B} {\bf 281} (1987) 310--364.

\bibitem{Monni:2011gb}
P.~F. Monni, T.~Gehrmann, and G.~Luisoni, {\it {Two-Loop Soft Corrections and Resummation of the Thrust Distribution in the Dijet Region}},  {\em JHEP} {\bf 08} (2011) 010, [\href{http://arxiv.org/abs/1105.4560}{{\tt arXiv:1105.4560}}].

\bibitem{Neill:2018uqw}
D.~Neill and W.~J. Waalewijn, {\it {Entropy of a Jet}},  {\em Phys. Rev. Lett.} {\bf 123} (2019), no.~14 142001, [\href{http://arxiv.org/abs/1811.01021}{{\tt arXiv:1811.01021}}].

\bibitem{Datta:2024hpn}
J.~Datta, A.~Deshpande, D.~E. Kharzeev, C.~J. Na{\"\i}m, and Z.~Tu, {\it {Entanglement as a Probe of Hadronization}},  {\em Phys. Rev. Lett.} {\bf 134} (2025), no.~11 111902, [\href{http://arxiv.org/abs/2410.22331}{{\tt arXiv:2410.22331}}].

\bibitem{Gong:2021bcp}
W.~Gong, G.~Parida, Z.~Tu, and R.~Venugopalan, {\it {Measurement of Bell-type inequalities and quantum entanglement from {\ensuremath{\Lambda}}-hyperon spin correlations at high energy colliders}},  {\em Phys. Rev. D} {\bf 106} (2022), no.~3 L031501, [\href{http://arxiv.org/abs/2107.13007}{{\tt arXiv:2107.13007}}].

\bibitem{Baumgart:2012ay}
M.~Baumgart and B.~Tweedie, {\it {A New Twist on Top Quark Spin Correlations}},  {\em JHEP} {\bf 03} (2013) 117, [\href{http://arxiv.org/abs/1212.4888}{{\tt arXiv:1212.4888}}].

\bibitem{CMS:2019nrx}
{\bf CMS} Collaboration, A.~M. Sirunyan et~al., {\it {Measurement of the top quark polarization and ${t\bar{t}}$ spin correlations using dilepton final states in proton-proton collisions at $\sqrt{s} =13 TeV$}},  {\em Phys. Rev. D} {\bf 100} (2019), no.~7 072002, [\href{http://arxiv.org/abs/1907.03729}{{\tt arXiv:1907.03729}}].

\bibitem{Clauser:1969ny}
J.~F. Clauser, M.~A. Horne, A.~Shimony, and R.~A. Holt, {\it {Proposed experiment to test local hidden variable theories}},  {\em Phys. Rev. Lett.} {\bf 23} (1969) 880--884.

\bibitem{Horodecki:1995nsk}
R.~Horodecki, P.~Horodecki, and M.~Horodecki, {\it {Violating Bell inequality by mixed spin- 1 2 states: necessary and sufficient condition }},  {\em Phys. Lett. A} {\bf 200} (1995), no.~5 340--344.

\bibitem{deFlorian:1997zj}
D.~de~Florian, M.~Stratmann, and W.~Vogelsang, {\it {QCD analysis of unpolarized and polarized Lambda baryon production in leading and next-to-leading order}},  {\em Phys. Rev. D} {\bf 57} (1998) 5811--5824, [\href{http://arxiv.org/abs/hep-ph/9711387}{{\tt hep-ph/9711387}}].

\bibitem{Soffer:1994ww}
J.~Soffer, {\it {Positivity constraints for spin dependent parton distributions}},  {\em Phys. Rev. Lett.} {\bf 74} (1995) 1292--1294, [\href{http://arxiv.org/abs/hep-ph/9409254}{{\tt hep-ph/9409254}}].

\bibitem{Vogelsang:1997ak}
W.~Vogelsang, {\it {Next-to-leading order evolution of transversity distributions and Soffer's inequality}},  {\em Phys. Rev. D} {\bf 57} (1998) 1886--1894, [\href{http://arxiv.org/abs/hep-ph/9706511}{{\tt hep-ph/9706511}}].

\bibitem{Burkardt:1993zh}
M.~Burkardt and R.~L. Jaffe, {\it {Polarized $q\to\Lambda$ fragmentation functions from $e^+e^-\to \Lambda+X^*$}},  {\em Phys. Rev. Lett.} {\bf 70} (1993) 2537--2540, [\href{http://arxiv.org/abs/hep-ph/9302232}{{\tt hep-ph/9302232}}].

\bibitem{Boros:1998kc}
C.~Boros and Z.-t. Liang, {\it {Spin content of Lambda and its longitudinal polarization in $e^+e^-$ annihilation at high-energies}},  {\em Phys. Rev. D} {\bf 57} (1998) 4491--4494, [\href{http://arxiv.org/abs/hep-ph/9803225}{{\tt hep-ph/9803225}}].

\bibitem{Ma:1998pd}
B.-Q. Ma and J.~Soffer, {\it {Quark flavor separation in Lambda Baryon fragmentation}},  {\em Phys. Rev. Lett.} {\bf 82} (1999) 2250--2253, [\href{http://arxiv.org/abs/hep-ph/9810517}{{\tt hep-ph/9810517}}].

\bibitem{Chen:2016iey}
K.-b. Chen, W.-h. Yang, Y.-j. Zhou, and Z.-t. Liang, {\it {Energy dependence of hadron polarization in $e^+e^-\to h\,X$ at high energies}},  {\em Phys. Rev. D} {\bf 95} (2017), no.~3 034009, [\href{http://arxiv.org/abs/1609.07001}{{\tt arXiv:1609.07001}}].

\bibitem{Accardi:2012qut}
A.~Accardi et~al., {\it {Electron Ion Collider: The Next QCD Frontier}: {Understanding the glue that binds us all}},  {\em Eur. Phys. J. A} {\bf 52} (2016), no.~9 268, [\href{http://arxiv.org/abs/1212.1701}{{\tt arXiv:1212.1701}}].

\bibitem{AbdulKhalek:2021gbh}
R.~Abdul~Khalek et~al., {\it {Science Requirements and Detector Concepts for the Electron-Ion Collider}: {EIC Yellow Report}},  {\em Nucl. Phys. A} {\bf 1026} (2022) 122447, [\href{http://arxiv.org/abs/2103.05419}{{\tt arXiv:2103.05419}}].

\bibitem{Anderle:2021wcy}
D.~P. Anderle et~al., {\it {Electron-ion collider in China}},  {\em Front. Phys. (Beijing)} {\bf 16} (2021), no.~6 64701, [\href{http://arxiv.org/abs/2102.09222}{{\tt arXiv:2102.09222}}].

\bibitem{Vogelsang:1996im}
W.~Vogelsang, {\it {The Spin dependent two loop splitting functions}},  {\em Nucl. Phys. B} {\bf 475} (1996) 47--72, [\href{http://arxiv.org/abs/hep-ph/9603366}{{\tt hep-ph/9603366}}].

\bibitem{Korchemsky:1987wg}
G.~P. Korchemsky and A.~V. Radyushkin, {\it {Renormalization of the Wilson Loops Beyond the Leading Order}},  {\em Nucl. Phys. B} {\bf 283} (1987) 342--364.

\bibitem{Moch:2004pa}
S.~Moch, J.~A.~M. Vermaseren, and A.~Vogt, {\it {The Three loop splitting functions in QCD: The Nonsinglet case}},  {\em Nucl. Phys. B} {\bf 688} (2004) 101--134, [\href{http://arxiv.org/abs/hep-ph/0403192}{{\tt hep-ph/0403192}}].

\bibitem{Moch:2005id}
S.~Moch, J.~A.~M. Vermaseren, and A.~Vogt, {\it {The Quark form-factor at higher orders}},  {\em JHEP} {\bf 08} (2005) 049, [\href{http://arxiv.org/abs/hep-ph/0507039}{{\tt hep-ph/0507039}}].

\bibitem{Moch:2005tm}
S.~Moch, J.~A.~M. Vermaseren, and A.~Vogt, {\it {Three-loop results for quark and gluon form-factors}},  {\em Phys. Lett. B} {\bf 625} (2005) 245--252, [\href{http://arxiv.org/abs/hep-ph/0508055}{{\tt hep-ph/0508055}}].

\bibitem{Kelley:2011ng}
R.~Kelley, M.~D. Schwartz, R.~M. Schabinger, and H.~X. Zhu, {\it {The two-loop hemisphere soft function}},  {\em Phys. Rev. D} {\bf 84} (2011) 045022, [\href{http://arxiv.org/abs/1105.3676}{{\tt arXiv:1105.3676}}].

\bibitem{Hornig:2011iu}
A.~Hornig, C.~Lee, I.~W. Stewart, J.~R. Walsh, and S.~Zuberi, {\it {Non-global Structure of the $\mathcal{O}({\alpha}^2_s)$ Dijet Soft Function}},  {\em JHEP} {\bf 08} (2011) 054, [\href{http://arxiv.org/abs/1105.4628}{{\tt arXiv:1105.4628}}]. [Erratum: JHEP 10, 101 (2017)].

\bibitem{Li:2011zp}
Y.~Li, S.~Mantry, and F.~Petriello, {\it {An Exclusive Soft Function for Drell-Yan at Next-to-Next-to-Leading Order}},  {\em Phys. Rev. D} {\bf 84} (2011) 094014, [\href{http://arxiv.org/abs/1105.5171}{{\tt arXiv:1105.5171}}].

\end{thebibliography}\endgroup

\end{document}